\documentclass{article}
\usepackage{epsfig} 
\usepackage{amsmath}
\usepackage{amsfonts}
\usepackage{graphicx}

\usepackage{a4}

\begin{document}

\title{Higgs- and Skyrme-Chern-Simons densities in all dimensions}
\author{
{\large D. H. Tchrakian}$^{\star \dagger}$ 
\\ 
\\
 \\ 
$^{\star}${\small School of Theoretical Physics, Dublin Institute for Advanced Studies,}\\
{\small 10 Burlington Road, Dublin 4, Ireland }\\
{$^\dagger$\small Department of Computer Science, Maynooth University, Maynooth, Ireland}}

\date{\today}
\newcommand{\dd}{\mbox{d}}
\newcommand{\tr}{\mbox{tr}}
\newcommand{\la}{\lambda}
\newcommand{\ka}{\kappa}
\newcommand{\f}{\phi}
\newcommand{\vf}{\varphi}
\newcommand{\vr}{\varrho}
\newcommand{\F}{\Phi}
\newcommand{\al}{\alpha}
\newcommand{\ga}{\gamma}
\newcommand{\Ga}{\Gamma}
\newcommand{\de}{\delta}
\newcommand{\si}{\sigma}
\newcommand{\Si}{\Sigma}
\newcommand{\ta}{\theta}
\newcommand{\Ta}{\Theta}
\newcommand{\bnabla}{\mbox{\boldmath $\nabla$}}
\newcommand{\om}{\omega}
\newcommand{\Om}{\Omega}
\newcommand{\bom}{\mbox{\boldmath $\omega$}}
\newcommand{\bOm}{\mbox{\boldmath $\Omega$}}
\newcommand{\bsi}{\mbox{\boldmath $\sigma$}}
\newcommand{\bchi}{\mbox{\boldmath $\chi$}}
\newcommand{\bal}{\mbox{\boldmath $\alpha$}}
\newcommand{\bpsi}{\mbox{\boldmath $\psi$}}
\newcommand{\brho}{\mbox{\boldmath $\varrho$}}
\newcommand{\beps}{\mbox{\boldmath $\varepsilon$}}
\newcommand{\bxi}{\mbox{\boldmath $\xi$}}
\newcommand{\bbeta}{\mbox{\boldmath $\beta$}}
\newcommand{\ee}{\end{equation}}
\newcommand{\eea}{\end{eqnarray}}
\newcommand{\be}{\begin{equation}}
\newcommand{\bea}{\begin{eqnarray}}

\newcommand{\ii}{\mbox{i}}
\newcommand{\e}{\mbox{e}}
\newcommand{\pa}{\partial}
\newcommand{\vep}{\varepsilon}
\newcommand{\bfph}{{\bf \phi}}
\newcommand{\lm}{\lambda}
\def\theequation{\arabic{equation}}
\renewcommand{\thefootnote}{\fnsymbol{footnote}}
\newcommand{\re}[1]{(\ref{#1})}
\newcommand{\R}{\mathbb R}
\newcommand{\N}{{\sf N\hspace*{-1.0ex}\rule{0.15ex}%
{1.3ex}\hspace*{1.0ex}}}
\newcommand{\Q}{{\sf Q\hspace*{-1.1ex}\rule{0.15ex}%
{1.5ex}\hspace*{1.1ex}}}
\newcommand{\C}{{\sf C\hspace*{-0.9ex}\rule{0.15ex}%
{1.3ex}\hspace*{0.9ex}}}
\newcommand{\eins}{1\hspace{-0.56ex}{\rm I}}
\renewcommand{\thefootnote}{\arabic{footnote}}

\maketitle


\bigskip

\begin{abstract}
Two types of new Chern-Simons (CS) densities, both defined in all odd and even dimensions, are proposed. These new CS densities
feature a scalar field interacting with the gauge field. In one case this is a Higgs scalar while in the other it is a Skyrme scalar. The motivation is
to study the effects of adding these new CS terms to a Lagrangian which supports static soliton solutions prior to their introduction.
\end{abstract}
\medskip
\medskip

\section{Introduction}
Chern-Simons (CS) densities play a prominent role~\cite{Jackiw:1985} in Yang-Mills (YM) field theory, both mathematically and physically.
Application of CS terms in $2+1$ dimensional YM field theory was introduced in \cite{Deser:1982vy,Deser:1981wh}, in the framework of the path integral formulation.
Because the dimension of the (quadratic) YM term differed from the dimension of the CS term by the first power of the dimensions of a mass, this was described as a
``topologically massive'' gauge theory~\footnote{The term ``topological'' is due to the quantisation of the dimensional constant of the theory.}. This is not the
context of the present report, which is concerned exclusively with the classical field theoretic aspects of Chern-Simons field theories.

Subsequent to the original work of \cite{Deser:1982vy}, CS terms were applied in classical field theories concerned with the construction of soliton solutions.
CS densities were used in Refs. \cite{Paul:1986ix,Hong:1990yh,Jackiw:1990aw}, to describe the dynamics of Abelian gauged-Higgs models in $2+1$ dimensions, which supported finite
energy soliton solutions. The salient feature of the CS dynamics in these cases was the introduction of electric charge and spin, in addition to the magnetic
vortex flux of the solitons. This was seen in \cite{Paul:1986ix} where the CS term was added to the Maxwell-Higgs Lagrangian, while in \cite{Hong:1990yh,Jackiw:1990aw}
the CS term was the sole source of the gauge field dynamics. (In the latter case the electrically charged solutions remained topologically stable, in addition to
having finite energy.) Still in $2+1$ dimensions and with Abelian gauge dynamics, analogues of the models in \cite{Hong:1990yh,Jackiw:1990aw}, with the complex
Higgs scalar being replaced by the $O(3)$ sigma model scalar, were studied in \cite{Ghosh:1995ze,Kimm:1995mi,Arthur:1996uu}.

The context of the present report is precisely that of \cite{Paul:1986ix,Hong:1990yh,Jackiw:1990aw} and \cite{Ghosh:1995ze,Kimm:1995mi,Arthur:1996uu}, namely that
of adding a CS term to the Lagrangian of a system that supports solitons prior to the introduction of the CS term, with a view to inquiring what effect this has on
the original solitons. Specifically, we aim to introduce a framework where (new) Chern-Simons densities can be defined in gauge theories in all (odd and even)
dimensional spectimes.

To date, what will be referred to below as Higgs-Chern-Simons (HCS) densities~\cite{Tchrakian:2010ar,Radu:2011zy},
were applied to Yang-Mills-Higgs (YMH) Lagrangians in $3+1$ dimensions and the resulting deformation of the solitons of the latter due to the influence
of these new HCS term(s) were investigated~\cite{Navarro-Lerida:2013pua,Navarro-Lerida:2014rwa}.
In the present report, we are not concerned with any such special applications. Rather, we aim to give a
general framework of constructing new Chern-Simons densities in all even and odd dimensions both for gauged-Higgs and gauged-Skyrme~\cite{Skyrme:1962vh} systems.

The (usual) Chern-Simons (CS) densities~\cite{Jackiw:1985} are defined in odd dimensions only.
This is because they descend from Chern-Pontryagin (CP) densities by $one$
dimension, the latter being defined as all possible traces of antisymmetrized products of the Yang-Mills (YM) curvature in even dimensions.
We will refer to these CS densities as the $usual$ CS densities.

The (usual) ${\cal N}$-th CP density ${\cal C}_{\rm CP}^{({\cal N})}$ in $2{\cal N}$ dimensions is by construction a $total\ divergence$
\be
\label{usualtt}
{\cal C}_{\rm CP}^{({\cal N})}=\bnabla\cdot\bOm^{({\cal N})}\equiv\pa_i\Om_i^{({\cal N})}
\ee
The usual CS density is then defined in $(2{\cal N}-1)$ (odd) dimensions as the $2{\cal N}$-th component of $\Om^{({\cal N})}_i$, $i.e.$
\be
\label{defusualCP}
\Om_{\rm CS}^{({\cal N})}\stackrel{\rm def.}=\Om^{({\cal N})}_{i=2{\cal N}}\,.
\ee
One way to construct CS densities in all (including even) dimensions is to start from such analogues of Chern-Pontryagin (CP) densities which are
defined in all (including odd) dimensions. This is what is proposed here. We consider two such analogues of CP densities, both employing
scalar fields in addition to the gauge fields. The central feature of both these analogue-CP densities is that they share the property of being
total-divergence with the $usual$ CP densities. Once these quantities are constructed, then the definintions of the new Chern-Simons densities
follow in the usual way $via$ the descent by one dimension indicated in \re{usualtt}-\re{defusualCP}.

The first of these analogue-CP densities are what one might call the Higgs--Chern-Pontryagin (HCP)
densities, which result from the dimensional reduction of the
$usual$ CP densities defined on direct product spaces $M^D\times K^N$, where $M^D$ is the $D$-dimensional residual space and $K^N$ is a
$N$-dimensional compact coset space. The most useful such spaces are the $N$-spheres $S^N$, to which we shall restrict our attention.
$D+N$ is of course $even$, but $N$ and $D$ individually can be either $odd$ or $even$. It turns out that after integrating out the coordinates
on $S^N$, the residual density on $M^D$ remains a total-divergence after the dimensional descent. These are the quantities that are
proposed as the analogue-CP densities, namely the Higgs--Chern-Pontryagin (HCP) densities. A fairly comprehensive sample of such densities is
given in Appendix {\bf A}, where a brief summary of the calculus of dimensional reduction employed is also presented.

We shall denote these HCP densities as ${\cal C}_{\rm HCP}^{({\cal N},D)}$ as the analogues of ${\cal C}_{\rm CP}^{({\cal N})}$,
with $D+N=2{\cal N}$.
Once we have these HCP densities (in all dimensions), the resulting (new) CS densities in $(D-1)$-dimensions can be constructed
$via$ the descent by one dimension in the usual way. We refer to these new CS densities as Higgs--Chern-Simons (HCS) densities.

The second of these analogue-CP densities is what one might call the Skyrme--Chern-Pontryagin (SCP) densities.
(By Skyrme~\cite{Skyrme:1962vh} systems we mean
$O(D+1)$ sigma models in $M^D$ defined in terms of the scalar field $\f^a$, $a=1,2,\dots,D+1$,
subject to the constraint $|\f^a|^2=1$.) In contrast to the
HCP densities mentioned in the previous paragraph, the SCP densities are not arrived at $via$ dimensional descent. They are rather the
topological charge densities~\cite{LMP} $\varrho^{(D,N)}$ of $SO(N)$ gauged Skyrme systems in (all even and odd) $D$ dimensions. Their analogy with the (usual) CP densities
${\cal C}_{\rm CP}^{({\cal N})}$ of YM systems is based on their shared property of being the topological charge densities of both systems, respectively. Specifically,
the crucial shared property is that both the $usual$ and the Skyrme--Chern-Pontryagin (SCP) densities are total-divergence.
The prescription for constructing the topological charge, $i.e.$ the SCP densities, is given in Appendix {\bf B}.
Again, a representative sample of these is listed in that Appendix. Again, the total-divergence property of the SCP densities enables the
definition of the corresponding Skyrme--Chern-Simons (SCS) densities in $(D-1)$-dimensions $via$ the usual descent \re{usualtt}-\re{defusualCP} by one dimension.

Appendices {\bf A} and {\bf B} serve the purpose of defining HCP and SCP densities, which are the building blocks in the constructions of the
Higgs--Chern-Simons (HCS) and Skyrme--Chern-Simons (SCS) densities, respectively. In the absence of induction formulas for the manifestly total divergence expressions
for the HCP and SCP densities, representative samples of these are listed, which happen to be of intrinsic interest in the construction of solitons.

The contents of Appendices {\bf A} and {\bf B} overlap substantially with parts of Refs. \cite{Tchrakian:2010ar} and \cite{LMP}, respectively.
They are given partly to make the account here more self-contained, but more importantly because in both cases the choices of gauge groups
made here are different from those made in \cite{Tchrakian:2010ar} and \cite{LMP}. Whereas previously in \cite{Tchrakian:2010ar,LMP}
the gauge groups were chosen such that topologically stable solitons in the $given$ dimensions exist, here, the corresponding choices
of gauge groups are made such that there exist topologically stable solitons in $other$ ($2$ steps lower) dimensions, where the solitons
to be deformed reside. The ultimate aim
for the application of the new Chern-Simons (CS) densities (both HCS and SCS) is to analyse the deformation of static stable solitons
due to the introduction of these new CS terms in the appropriate Lagrangians.

In Section ${\bf 2}$, we give the prescription for constructing the Higgs--Chern-Simons (HCS) densities, present several examples of possible
physical interest, and discuss some of their properties with a view to application. A similar
account of the Skyrme--Chern-Simons (SCS) densities is presented in Section ${\bf 3}$. Sections ${\bf 2}$ and ${\bf 3}$ rely, respectively, on the
corresponding definitions of the Higgs--Chern-Pontryagin (HCP) and the Skyrme--Chern-Pontryagin (SCP) densities presented in
Appendices {\bf A} and {\bf B}. In Section ${\bf 4}$, a summary and a brief discussion of possible applications is given.

\section{Higgs-Chern-Simons (HCS) densities}
Before presenting the Higgs-Chern-Simons (HCS) densities, it should be stressed that the context in which the application of these HCS terms
is envisaged, is their effect on the soliton solutions of a Yang-Mills--Higgs (YMH) system before the introduction of the HCS term to its
Lagrangian. This circumstance puts a strong restriction on the multiplet structure of the Higgs field(s) appearing in the system.
The point is that whatever is the Higgs multiplet of the original YMH model that is consistent with achieving
finite energy
asymptotics, the same Higgs multiplet must be the one also defining the HCS term added to the YMH Lagrangian.

In Appendix {\bf A}, we have listed the HCP densities from which the HCS terms will be derived here. The Higgs fields $\F$ considered are
$n\times n$ and $n\times n\oplus n\times n$ multiplets, having eschewed the option of Higgs fields $\f$ in the fundamental
representation of $SU(n)$ given by \re{fund12}-\re{fund1}. This is because the ``physical'' aim is to study the effect
of adding HCS term(s) to the YMH Lagrangian on the (finite energy) solutions of the YMH system, while insisting that the HCS-deformed
solution also supports finite energy. Thus if the Higgs field $\F$ is employed in the original YMH Lagrangian, and we employ $\f$ to define the HCS term,
then we would have to satisfy the two $finite\ energy$ conditions
\be
\label{simult}
\lim_{r\to\infty}\,D_i\F=0 \quad{\rm and}\quad\lim_{r\to\infty}\,D_i\f=0
\ee
simultaneously, which is clearly impossible.

Thus requiring that the solutions of the HCS-extended YMH model (the YMH-HCS model)
have finite energy requires that $either$ the Higgs field $\F$, $or$ $\f$, must be employed exclusively.
Further requiring that the soliton of the YMH
system be topologically stable~\footnote{Except in $3+1$ dimensions, where there exist $sphalerons$ of the YMH system with a $fundamental$
Higgs, in other dimensions not even such unstable solutions are known.} disqualifies the use of the Higgs multiplet $\f$.
For this reason fundamental Higgs fields $\f$ are excluded from our considerations.

\medskip
\noindent
{\bf The Higgs-Chern-Simons} (HCS) densities are extracted from the Higgs-Chern-Pontryagin (HCP) densities in the usual way $via$ the descent
\re{usualtt}-\re{defusualCP} by one dimension. The HCP
are the dimensional descendants of the CP, the details of which descents are presented in Appendix {\bf A}.

The crucial property that HCP densities ${\cal C}_{\rm HCP}^{({\cal N},D)}$ in (residual) $D$ dimensions share with the ${\cal N}$-th CP
density ${\cal C}_{\rm CP}^{({\cal N})}$ in $2{\cal N}$ dimensions is that like the latter, they are $total\ divergence$, which we denote as
\be
\label{totdiv}
{\cal C}_{\rm HCP}^{({\cal N},D)}=\bnabla\cdot\bOm^{({\cal N},D)}\,.
\ee
Explicit expressions of $\bOm^{({\cal N},D)}\equiv\Om_i^{({\cal N},D)}$ ($i=1,2,\dots,D$) are given by fo 
${\cal N}=3,4$ for several relevant examples
by \re{Om35}-\re{Om44}, in Appendix {\bf A}.

The definition of the Higgs-Chern-Simons (HCS) density pertaining to a given Higgs-Chern-Pontryagin (HCP) density
$\Om_i^{({\cal N},D)}$ is that it is the $D$-th (or any other) component of $\Om_{i}^{({\cal N},D)}$, $i.e.$
\be
\label{defn}
\Om_{\rm HCS}^{({\cal N},D-1)}\stackrel{\rm def.}=\Om_{i=D}^{({\cal N},D)}\,,
\ee
like in \re{usualtt}-\re{defusualCP}. The density $\Om_{\rm HCS}^{({\cal N},D-1)}$ is a scalar in the relevant $(D-1)$-dimensional space(-time) with coordinates
$x_{\mu}$, ($\mu=1,2,\dots,(D-1)$), and is a functional of $[A_{\mu},\F]$.

For the examples of HCP densities listed in Appendix {\bf A}, namely \re{Om35}-\re{Om34} and \re{Om47}-\re{Om44},
the corresponding HCS terms can be read off immediately. Those in $3+1$ and $2+1$
dimensional space(-times) resulting from the ${\cal N}=3$ HCP densities are, respectively
\bea
\Omega^{(3,5-1)}_{\rm HCS}&=&\vep_{\mu\nu\rho\si}\,\mbox{Tr}\ F_{\mu\nu}\,F_{\rho\si}\,\F\label{HCS35}\\
\Omega^{(3,4-1)}_{\rm HCS}&=&\vep_{\mu\nu\la}\mbox{Tr}\,\hat\Gamma^{(n)}\bigg[-2\eta^2A_{\la}\left(F_{\mu\nu}
-\frac23\,A_{\mu}A_{\nu}\right)+\left(\F\, D_{\la}\F-D_{\la}\F\, \F\right)F_{\mu\nu}\bigg]\label{HCS34}
\eea
and those in dimensional space(-times) $5+1$, $4+1$, $3+1$ and $2+1$ resulting from the ${\cal N}=4$ HCP densities are,
\bea
\Omega^{(4,7-1)}_{\rm HCS}&=&\vep_{\mu\nu\rho\si\tau\la}\,\mbox{Tr}\ F_{\mu\nu}\,F_{\rho\si}\,F_{\tau\la}\,\F\label{HCS47}\\
\Omega^{(4,6-1)}_{\rm HCS}&=&\vep_{\mu\nu\rho\si\la}\,\mbox{Tr}\,\hat\Gamma^{(n)}\,A_{\la}\bigg[\left(F_{\mu\nu}F_{\rho\si}-
F_{\mu\nu}A_{\rho}A_{\si}+\frac25A_{\mu}A_{\nu}A_{\rho}A_{\si}\right)+\nonumber\\
&&\hspace{40mm}+D_{\la}\F\left(\F F_{\mu\nu}F_{\rho\si}+F_{\mu\nu}\F F_{\rho\si}+F_{\mu\nu}F_{\rho\si}\F\right)\bigg]\label{HCS46}\\
\Omega^{(4,5-1)}_{\rm HCS}&=&\vep_{\mu\nu\rho\si}\,\mbox{Tr}\bigg[
\F\left(\eta^2\,F_{\mu\nu}F_{\rho\si}+\frac29\,\F^2\,F_{\mu\nu}F_{\rho\si}+\frac19\,F_{\mu\nu}\F^2F_{\rho\si}\right)
\nonumber\\
&&\qquad\qquad\qquad-\frac29\left(\F D_{\mu}\F D_{\nu}\F-D_{\mu}\F\F D_{\nu}\F+D_{\mu}\F D_{\nu}\F\F\right)F_{\rho\si}\bigg]\label{HCS45}\\
\Omega^{(4,4-1)}_{\rm HCS}&=&\vep_{\mu\nu\la}\,\mbox{Tr}\,\hat\Gamma^{(n)}\,\bigg\{6\eta^4\,A_{\la}\left(F_{\mu\nu}-\frac23\,A_{\mu}\,A_{\nu}\right)
-6\,\eta^2\left(\F\,D_{\la}\F-D_{\la}\F\,\F\right)\,F_{\mu\nu}\nonumber\\
&&\hspace{20mm}+\bigg[\left(\F^2\,D_{\la}\F\,\F-\F\,D_{\la}\F\,\F^2\right)-2\left(\F^3\,D_{\la}\F-D_{\la}\F\,\F^3\right)\bigg]F_{\mu\nu}
\bigg\}\label{HCS44}
\eea
where the dimensionful constsnt $\eta$ is the inverse of the radius of the sphere over which the dimensional descent resulting in the HCP densities is performed.

It transpires that in all odd dimensional spacetimes the HCS densities,
$e.g.$ \re{HCS34}, \re{HCS46} and \re{HCS44}, the leading term consists of the $usual$ CS
density for the $SU(n)\times SU(n)$ gauge field which is $gauge\ variant$, and a Higgs dependant part which is
$gauge\ invariant$. The gauge transformation properties of these HCS densities
are therefore precisely the same as those of the $usual$ CS densities in the given (odd) dimensions.

By contrast, in all even dimensional spacetimes the HCS densities, $e.g.$ \re{HCS35}, \re{HCS47} and \re{HCS45}, are
$gauge\ invariant$ and are expressed by both the $SU(n)$ gauge field and the
corresponding algebra valued Higgs field. Obviously, these do not have $usual$ CS analogues and are in that sense entirely new.
This procedure can be repeated for all ${\cal N}\ge 5$ yielding a tower of HCS densities in a spacetime of any given dimension.

As in the $2+1$ dimensional applications~\cite{Paul:1986ix,Hong:1990yh,Jackiw:1990aw} , the salient feature of the
Chern-Simons dynamics is the introduction of the electric component $A_0$ of the gauge connection
and hence the electric charge. Of course, in odd dimensional spacetimes one has the $usual$
Chern-Simons (CS) density so the application of HCS terms is not indispensible. On the other hand
it is conceivable that HCS dynamics may lead to new qualitative effects. Certainly in even dimensional
spacetimes one has no other choice than to employ HCS densities, and in the
$3+1$ dimensional cases studied~\cite{Navarro-Lerida:2013pua,Navarro-Lerida:2014rwa} not only does this enable the electrical
charging of the soliton, but the effects of the two HCS densities
\re{HCS35} and \re{HCS45} employed there result in qualitatively different effects.

We end our account of the application of HCS, with attention to the case when the gauge group is Abelian. This
is possible only for HCP densities in even $D$, $i.e.$ in odd
dimensional spacetimes in which the HCS density may play a dynamical role. In particular in $2+1$ dimensions, solitons (vortices) of the YMH (Maxwell-Higgs) system
exist only for $U(1)$. But as explained at the end of Section {\bf A.3}, Abelian HCP densities in $D=4p\ ,\quad p=1,2,\dots$ vanish
but are nonvanishing in $D=4p+2\ ,\quad p=0,1,2,\dots$, so for an Abelian Higgs system in $2+1$ dimensions there exists no HCS density since this would result from the corresponding
$D=4$ HCP density, which is absent.
Of course in that case, one still has
the $usual$ CS density, which would have been present as the leading term even if the HCS density had existed.

Thus one can define Abelian HCS densities only in $4+1$, $6+1$, $etc.$
dimensions. These can be read off \re{calN=2p+2,D=4p+2}. For example, the Abelian HCS density in $4+1$ dimensions is
\be
\label{abelHCS5}
\tilde{\cal C}_{\rm HCP}^{(4,6)}
=\vep_{\mu\nu\rho\si\la}\left[f_{\mu\nu}f_{\rho\si}\left(\eta^2\,a_{\la}+i\vf^*D_{\la}\vf\right)\right]\,.
\ee
It is of course not possible to construct (regular) solitons of a system featuring a complex scalar field in these dimensions, but \re{abelHCS5}
can be useful in studying black holes since in that case it is not necessary that the solution be regular at the origin.

\section{Skyrme-Chern-Simons (SCS) densities}
The definition of the Skyrme-CS (SCS) density in $D-1$ dimensions proceeds exactly as for the usual CS and the Higgs-CS (HCS)
densities stated in \re{defusualCP} and \re{defn}, respectively. This
involves the one-step descent of the corresponding Skyrme--Chern-Pontryagin (SCP) densities in $D$ dimensions, which are presented in Section {\bf B.2} of Appendix {\bf B}.

Unlike in the case of the Higgs-CS (HCS) densities in odd dimensions, where the leading term was the usual CS density in those dimensions, the systematic derivation of
the the Skyrme-CS (SCS) density does not feature such a term. In the former case, the HCS, the mechanism responsible for the appearance of the usual
CS term in the HCS density results from the presence of the usual Chern-Pontryagin (CP) term in the definition of the Higgs-CP density, the latter
being a dimensional descendent of a CP density.
By contrast, the construction of the Skyrme--Chern-Pontryagin (SCP) density does not involve the dimensional reduction of a (usual) CP density
in higher dimensions, so that no residual CP term appears there.

The SCP densities in $D$ dimensions used in the construction of the SCS densities in $D-1$ dimensions
are ${\cal C}_{\rm SCP}^{(D,N)}$ and $\hat{\cal C}_{\rm SCP}^{(D,N)}$, defined by \re{SCP} and \re{SCPe} for odd- and even-$D$ respectively.
There, we have modified the definition, \re{SCPe}, of the SCP density in even $D$ by hand,
such that it include the usual CP density.

The definitions of the corresponding Skyrme-CS (SCS) densities then follow systematically.
These SCS densities, $\Om_{\rm SCS}^{(D-1,N)}$ and $\hat\Om_{\rm SCS}^{(D-1,N)}$, in $D-1$ dimensional spacetime, are defined $via$
the one-step descent as
\bea
\Om_{\rm SCS}^{(D-1,N)}&\stackrel{\rm def.}=&\left(\om^{(D)}_{i_D=D}+\Om_{i_D=D}^{(D,N)}\right)\ ,\quad D-1 \quad{\rm even}\label{defnSodd}\\
\hat\Om_{\rm SCS}^{(D-1,N)}&\stackrel{\rm def.}=&\left(\om^{(D)}_{i_D=D}+\Om_{i_D=D}^{(D,N)}\right)+\la\,\Om^{({\cal N})}_{i=2{\cal N}}
\ ,\quad D-1 \quad{\rm odd}\label{defnSeven}
\eea
where $\Om^{({\cal N})}_{i=2{\cal N}}=\Om^{({\cal N})}_{\rm CS}$ in \re{defnSeven} is the familiar (usual) CS density
for the $SO(N)$ gauge field in $2{\cal N}-1$ dimensions appearing in \re{usualtt}. The real parameter $\la$ appearing in \re{defnSeven} is the coupling strength of the usual
CS density $\Om^{({\cal N})}_{\rm CS}$ in this (odd) dimension.

In \re{defnSodd}-\re{defnSeven}, the quantities $\om^{(D)}_{i_D=D}$ and $\Om_{i_D=D}^{(D,N)}$ are scalar densities in $D-1$ dimensional
spacetime, whose coordinates $x_{\mu}$ are labelled by $\mu,\nu,\dots$, with $\mu,\nu=1,2,\dots,D-1$. The quantities $\Om_{i_D=D}^{(D,N)}$ in particular encode
the integer $N$, pertaining to the gauge group $SO(N)$, $2\le N\le D-1$.

We do not have a closed form expression for the arbitrary case of $\Om_{i_D=D}^{(D,N)}$, but a general expression for
the term $\om^{(D)}_{i_D=D}$ in \re{defnSodd}-\re{defnSeven} can be read off \re{wN3} and \re{denote} as
\be
\label{ommink}
\om^{(D)}_{i_D=D}=\vep_{\mu_1\mu_2\dots\mu_{D-1}}\,\left(\sin^{D-1}f^{(1)}\pa_{i_{\mu1}}f^{(1)}\right)\left(\sin^{D-2}f^{(2)}\pa_{i_{\mu_2}}f^{(2)}\right)
\dots\left(\sin f\pa_{i_{\mu_{D-1}}}f^{(D-1)}\right)g\,.
\ee
Before we list the chosen examples of SCS densities, we make some simplifications and refinements of our notation. To avoid writing $\om^{(D)}_{i_D=D}$ repeatedly, we
instead write
\be
\label{simpom}
\om^{(D)}\equiv\,\om^{(D)}_{i_D=D}\,.
\ee
Also, since we are concerned only with $SO(N)$ gauge groups here, we find it helpful to
 modify the notation used in \re{defusualCP} for the usual ${\cal N}$-th CP density $\Om_{\rm CS}^{({\cal N})}$, to take account of the $SO(N)$ gauge group here as follows
\be
\label{simpOm}
\Om_{\rm CS}^{({\cal N})}\to\Om_{\rm CS}^{({\cal N},N)}\,.
\ee
We finally list the SCS densities that follow from the definitions of the SCP densities listed in Section {\bf B.3}.
These are displayed in groups according to the dimensionality of the space(-time), $D-1$, and gauge group $SO(N)$ in each case.

\medskip
\noindent
{\bf In $2+1$ dimensions} the SCS density \re{defnSeven} results from the descents carried out on the $D=4$ SCP densities \re{defnSeven},
namely from \re{SCP42b}, \re{SCP43b} and \re{SCP44b}, with gauge groups $SO(2)$, $SO(3)$ and $SO(4)$ respectively. These are
\bea
\hat\Om_{\rm SCS}^{(4-1,2)}&=&\left(\om^{(4)}+\vep_{\mu\nu\la} \vep^{ABC}
A_{\la}\,\pa_{\mu}\f^{{A}}\,\pa_{\nu}\f^{{B}}\,\f^{{C}}\right)+\la\,\Om^{(2,2)}_{\rm CS}\label{SCS42}\\
\hat\Om_{\rm SCS}^{(4-1,3)}&=&\left(\om^{(4)}+\vep_{\mu\nu\la} \vep^{AB}
A_{\la}^{\al}\f^{\al}\,\pa_{\mu}\f^{A}\,\pa_{\nu}\f^{B}\right)+\la\,\Om^{(2,3)}_{\rm CS}\label{SCS43}\\
\hat\Om_{\rm SCS}^{(4-1,4)}&=&\om^{(4)}+\frac32\vep_{\mu\nu\la}\vep^{\alpha \alpha'\beta\beta'}\Bigg(
\phi^5 (1-\frac{1}{3} (\phi^5)^2) A_{\la}^{\al\al'}
\left(\pa_{\mu} A_{\nu}^{\beta\beta'}+\frac{2}{3} (A_{\mu} A_{\nu} )^{\beta \beta'}\right)- \nonumber\\
&&\hspace{20mm}+3\,\f^5
\bigg[F_{\mu\nu}^{\al\al'}\f^{\beta}D_{\la}\f^{\beta'}-\pa_{\mu}\left[A_{\nu}^{\al\al'}\f^{\beta}\left(2\pa_{\la}\f^{\beta'}+A_{\la}\f^{\beta'}\right)\right]\bigg]\Bigg)+\la\,\Om^{(2,4)}_{\rm CS}
\label{SCS44}
\eea
The (usual) $SO(2)$, $SO(3)$ and $SO(4)$ CS densities $\Om^{(2,2)}_{\rm CS}$, $\Om^{(2,3)}_{\rm CS}$ and $\Om^{(2,4)}_{\rm CS}$ in \re{SCS42}, \re{SCS43} and \re{SCS44} are given by
\bea
\Om^{(2,2)}_{\rm CS}&=&\vep_{\mu\nu\la}\,A_{\la}\,\pa_{\mu}A_{\nu}\label{CSabel}\\
\Om^{(2,3)}_{\rm CS}&=&\vep_{\mu\nu\la}\,A_{\la}^{\al}
\left(\pa_{\mu} A_{\nu}^{\al}+\frac{2}{3} (A_{\mu} A_{\nu} )^{\al}\right)\label{CS21}\\
\Om^{(2,4)}_{\rm CS}&=&\vep_{\mu\nu\la}\,\vep_{\al\al'\beta\beta'}A_{\la}^{\al\al'}
\left(\pa_{\mu} A_{\nu}^{\beta\beta'}+\frac{2}{3} (A_{\mu} A_{\nu} )^{\beta \beta'}\right)\label{CS2}
\eea

\medskip
\noindent
{\bf In $3+1$ dimensions} the SCS density results from the one-step descents carried out on the $D=5$ SCP densities \re{defnSodd},
namely from \re{SCP52b} and \re{SCP53b}, with gauge groups $SO(2)$ and $SO(3)$ respectively. These are
\bea
\Om_{\rm SCS}^{(5-1,2)}&=&\om^{(5)}+2\,\vep_{\mu\nu\rho\si}\vep^{{A}{B}{C}{D}}\,A_{\mu}\,
\pa_{\nu}\f^{{A}}\pa_{\rho}\f^{{B}}\pa_{\si}\f^{{C}}\f^{{D}}\label{SCS52}\\
\Om_{\rm SCS}^{(5-1,3)}&=&\om^{(5)}+\vep_{\mu\nu\rho\si}\vep^{{A}{B}{C}}\,A_{\mu}^{\al}\,\f^{\al}
\pa_{\nu}\f^{{A}}\pa_{\rho}\f^{{B}}\pa_{\si}\f^{{C}}\,.\label{SCS53}
\eea

It may be interesting to note that the Skyrme-CS (SCS) densities in both even and odd dimensional space(-times) are gauge $variant$ quantities in contrast to the
Higgs-CS (HCS) densities, which in even dimensional spacetimes are $gauge\ invariant$.


\section{Summary and Discussion}
This report is concerned with the definitions of two (new) types of Chern-Simons (CS) densities, to which we have referred as Higgs--Chern-Simons (HCS)
and Skyrme--Chern-Simons (SCS) densities. They are both defined in all (odd and even) dimensions unlike the usual CS densities which are defined
in odd dimensions only. Both depend on a given Yang-Mills (YM), or Maxwell, curvature and its connection. In addition to the gauge field, the HCS depend on a Higgs scalar,
while the SCS depend on a Skyrme scalar.

Just as the definition of the usual CS density relies on the CP density being total divergence, here too
the definitions of both the HCS and SCS densities in $D-1$ dimensional space(-time) rely on the fact that the respective HCP and SCP densities in $D$
dimensions are also $total\ divergence$. The HCP and SCP densities are just the topological charge densities
of gauged-Higgs and gauged-Skyrme systems in $D$ dimensions. The topological charge densities of the gauged-Higgs systems result from the dimensional reduction
of the CP density in the bulk over spheres, presented in Appendix {\bf A}. The topological charge densities of the gauged-Skyrme systems are constructed
directly by specifying a gauging prescription, given in Appendix {\bf B}.

They (the HCS and SCS) are both arrived at $via$ a one-step descent of the respective Chern-Pontryagin (CP) analogue, namely from the Higgs--Chern-Pontryagin (HCP)
and the Skyrme--Chern-Pontryagin (SCP) in $D$ dimensions, respectively. The resulting HCS and SCS are defined in $D-1$ dimensions -- the spacetime. In the context here,
it is envisaged that the models in question whose Lagrangians are augmented with the HCS should support finite energy solutions carrying magnetic charge in $\R^{D-2}$ prior the introduction of the HCS,
while the models whose Lagrangians are augmented with SCS terms should support topologically stable solitons carrying baryon charge in $\R^{D-2}$.

While the provenance of both the HCS and SCS densities is conceptually the same as that of the $usual$ CS density, unlike the latter
their definitions are not restricted to odd dimensional spacetimes. Many other properties of the HCS and SCS are quite contrasting, as
described below.

\begin{itemize}
\item
{\bf Gauge group}

The usual CS density can be defined for any gauge group, namely the gauge group of the CP density from which it is descended by one step.
By contrast the gauge groups of both the HCS and the SCS
introduced here, are much more restrictive. In both cases these restrictions follow from the restrictions on the gauge groups of both the HCP and the SCP densities
from which the former are descended.

\begin{itemize}
\item
{\bf Gauge group of the HCS density}
\begin{itemize}
\item
In even dimensional spacetime the gauge group is $SU(n)$, the choice of $n$ being determined by the choice of the dimension $2{\cal N}$ of the bulk space
prior to the dimensional reduction yielding the pertinent HCP density.
\item
In odd dimensional spacetime the gauge group is $SU(n)\times SU(n)$, $n$ being again determined by the choice of $2{\cal N}$ of the bulk.
\item
$Abelian\ gauge\ group$: This is possible only in odd dimensional spacetimes. It is further restricted to $4p+1$ dimensions, for $p=0,1,2,\dots$
\end{itemize}
\item
{\bf Gauge group of the SCS density}

In $D-1$ dimensional spacetime, the gauge group can be $SO(N)$ for all possible $N$ subject to $2\le N\le D$. (This restriction results from our prescription of
gauging the $O(D+1)$ Skyrme model in $D$ dimensions.)
\end{itemize}
\item
{\bf Scalar multiplets}
\begin{itemize}
\item
{\bf Higgs multiplet}

In $D-1$ dimensional spacetime, the choice of the Higgs multiplet is made subject to two criteria. $i)$ the gauge group must be large enough for the Higgs-CS
densities to be nonvanishing, and $ii)$ that the static solutions of the system on $\R^{D-2}$,
prior to the introduction of the HCS terms to the YMH Lagrangian, have at least finite energy if not also topological stability. Subject to these criteria,

\begin{itemize}
\item
In even dimensional spacetime the Higgs field takes its values in the $adjoint\ representation$ of the gauge group $SU(n)$.
\item
In odd dimensional spacetime the Higgs field takes its values in $su(n)\oplus su(n)$ transforming under $SU(n)\times SU(n)$. 
\item
$Abelian\ gauge\ group$: In this case the Higgs field is a complex scalar.
\end{itemize}
In all cases both the HCS term(s) and the YMH Lagrangian prior to their introduction, must be described by the same Higgs multiplet $\F$. Finite energy
conditions cannot be satisfied for a given gauge group, for two distict Higgs multiplets, $cf$ Eqn. \re{simult}.
\item
{\bf Skyrme multiplet}

In $D-1$ dimensional spacetime, whether $D$ is even or odd, and irrespective of the choice of gauge group $SO(N)$, the full Lagrangian of the system is described by
two distinct Skyrme scalars, one defining the SCS term and the other defining the model prior to the introduction of the SCS term.
\begin{itemize}
\item
For any given $D$, the Skyrme-CS (SCS) in $D-1$ dimensions is defined in terms of the $O(D+1)$ sigma model field $\f^a$, $a=1,2,\dots,D+1$, subject to $|\f^a|^2=1$.
\item
For any given $D$, the system prior to the introduction of the SCS term to the Lagrangian is the $O(D-1)$ sigma model field $\ta^{\bar a}$, $\bar a=1,2,\dots,D-1$, subject to $|\ta^{\bar a}|^2=1$.
This system supports static, finite energy and be topologically stable solutions in $\R^{D-2}$, as required.

\end{itemize}

 The two Skyrme scalars see each other only through the chosen $SO(N)$ gauge field, with $N$ subject to $2\le N\le D-1$.
\end{itemize}
\item
{\bf Gauge transformations}
\begin{itemize}
\item
Gauge transformations of the HCS
\begin{itemize}
\item
In even dimensional spacetime the HCS density is $gauge\ invariant$, rendering these irrelevant for application to ``topologically massive'' gauge theory.
\item
In odd dimensional spacetimes the HCS density is $gauge\ variant$. It consists of a leading term which is
the $usual$ CS density in that dimension, plus a Higgs dependent part which is $gauge\ invariant$.
\begin{itemize}
\item
The (real) dimensional coefficient of the $usual$ CS component of the HCS is fixed by the dimensional descent over the spheres.
\item
Its gauge transformation properties therefore are precisely those of the $usual$ CS density~\footnote{The gauge variation consists of two terms in each dimension.
One is a total divergence and the other has the form of a winding number density. For the integral of the total divergence term to vansih, and for the integral
of the winding number density term to yield an integer, the appropriate asymptotic behaviour of the gauge connection must be assumed. This will differ from the
asymptotic behaviour required of the usual YM-CS theories because of the presence of the Higgs field.} so subject to the additional requirements due to the presence
of the Higgs scalar, it is potentaially applicable to ``topologically massive'' gauge theories.
\end{itemize}
\end{itemize}
\item
Gauge transformations of the SCS
\begin{itemize}
\item
In both even and odd dimensional spacetimes the SCS density is $gauge\ variant$, since
both $\om^{(D)}=\om^{(D)}_{i_D=D}$ and $\Om_{i_D=D}^{(D,N)}$ featuring in its definitions\re{defnSodd} and \re{defnSeven},
are each separately $gauge\ variant$.
\item
Because of the presence of the term $\om^{(D)}$ in the definitions of all SCS densities, the gauge variations of these cannot be cast in a geometrically
convenient form. Hence, SCS densities are not candidates for CS terms in ``topologically massive'' gauge theories.
\end{itemize}
\end{itemize}
\end{itemize}

\medskip
\noindent
{\bf In the context of applications}, perhaps the most important feature of both the Higgs and Skyrme cases is that they
enable the electrical charging of the soliton, and in certain cases also allow the soliton to spin. While we are not actively concerned here with applications,
it is nonetheless in order to discuss the possibile consequences of adding the HCS and SCS terms to various Lagrangians. Specifically it is interesting to inquire what
influence these new CS terms can have on the solitons of the system which were supported prior to their introduction to the Lagrangian, $e.g.$
the electrical charging of the soliton, and, changes in its mass and spin.

There are remarkable differences in the properties of HCS and SCS terms when applied to gauged Higgs and gauged Skyrme Lagrangians. The gauged Higgs system, $i.e.$
the Yang-Mills--Higgs (YMH) system, when subjected to the influence of HCS terms can be labelled as YMH-HCS, where only one Higgs field is employed. By contrast,
the gauged Skyrme system, say YMS$_1$, prior to the introduction the of the SCS$_2$ term may be labelled as YMS$_1$-SCS$_2$, in a notation emphasising the distinction between
the two Skyrme scalars involved.

We list some of these differences below.
\begin{itemize}
\item
In a given dimension, the YMH-HCS Lagrangian can (in principle) feature an infinite tower of HCS terms since these result from dimensional reduction, with the choice
of codimension being unlimited.
\item
In a given dimension, both the gauged-Skyrme Lagrangian YMS$_1$ and the SCS term SCS$_2$ are uniquely fixed by the choice of gauge group $SO(N)$ allowed by the gauging prescription.
\item
In the case of a YMH system, a given choice of gauge group $SU(n)$, or $SU(n)\times SU(n)$, which is adequate to sustain nonvanishing HCS terms, while it will be adequate
to support static finite energy solutions in $\R^{D-2}$, it will not in general~\footnote{The possiblity of assuring stability by arranging to have self-dual
solutions prior to the introduction of the HCS terms is absent, except in $3$ dimensions. This is because in greater than $3$ dimensions, the self-duality equations of YMH systems are
overdetermined~\cite{Tchrakian:2010ar}.} render these topologically stable. Only self-dual solutions can be stable, but even then finding a simple
Ansatz for higher gauge groups is not a simple task.
\item
The static Skyrmion in $\R^{D-2}$ of the gauged $O(D-1)$ sigma model YMS$_1$ is topologically stable for any allowed gauge group $SO(N)$. The
SCS$_2$ term introduced to the Lagrangian is defined on the other hand, in terms of a different Skyrme scalar, namely that of the $O(D+1)$ sigma model. In the given dimension $\R^{D-2}$,
there exists no topological density in terms of the $O(D+1)$ Skyrme scalar. As a result, the calssical stability of the deformed soliton is compromised.
\end{itemize}

To date there has only been one application~\cite{Navarro-Lerida:2013pua,Navarro-Lerida:2014rwa} of new CS dynamics,
namely in the $SO(5)$ and $SU(3)$ gauged YMH systems, in $3+1$ dimensional spacetime. The YMH Lagrangian was augmented with the first two members
in the hierarchy of Higgs-CS terms. The qualitative and quantitative properties of the solutions resulting from these two members of the HCS tower were effectively different.
Before adding the HCS terms, the YMH system supported electrically charged Julia-Zee~\cite{Julia:1975ff} (JZ) type static solitons in $\R^3$. As expected, the energies of
the electrically charged JZ type solutions are {\bf larger} than those of the electrically neutral ones. But when the HCS term(s) were introduced,
it was found that the energies of the electrically charged solutions were {\bf smaller} than those of the electrically neutral ones, in some regions of the parameter space.

Of course whether this effect persists in other other systems must be tested.
The most interesting future investigations are those empolying SCS terms to deform gauged Skyrmions. These involve adding the SCS
terms $\Om_{\rm SCS}^{(D-1,N)}$ and $\hat\Om_{\rm SCS}^{(D-1,N)}$ given by \re{defnSodd}-\re{defnSeven} defined in terms of the $O(D+1)$ Skyrme scalar to the Lagrangian of the $O(D-1)$ Skyrme model,
in $D-1$ dimensional spacetime.

When studying static solutions, as is the case envisaged here, the definitions of these SCS terms simplifies.
This simplification is of crucial importance in practice. Concerning $\Om_{i_D=D}^{(D,N)}$ in \re{defnSodd}-\re{defnSeven},
while we have no closed form expression for the arbitrary case, their construction in any given case is uniquely determined by the presecription of gauging the relevant sigma model.
Examples of these are seen in \re{SCS42}-\re{SCS53} above. The definitions of $\om^{(D)}\equiv\om^{(D)}_{i_D=D}$ by contrast can be given in closed form in the general case but, this definition is
not unique. This non-uniqueness however is of no consequence when restricting attention to the static limit. Consider for illustration the example \re{ommink} that we had privleged above,
\be
\nonumber
\om^{(D)}=\vep_{\mu_1\mu_2\dots\mu_{D-1}}\,\left(\sin^{D-1}f^{(1)}\pa_{i_{\mu1}}f^{(1)}\right)\left(\sin^{D-2}f^{(2)}\pa_{i_{\mu_2}}f^{(2)}\right)
\dots\left(\sin f\pa_{i_{\mu_{D-1}}}f^{(D-1)}\right)g\,.
\ee
It is obvious that this $\om^{(D)}$ {\bf vanishes} in the static limit.

As a result, \re{defnSodd}-\re{defnSeven} simplify to
\bea
\Om_{\rm SCS}^{(D-1,N)}&\stackrel{\rm def.}=&\Om_{i_D=D}^{(D,N)}\ ,\quad D-1 \quad{\rm even}\nonumber\\
\hat\Om_{\rm SCS}^{(D-1,N)}&\stackrel{\rm def.}=&\Om_{i_D=D}^{(D,N)}+\la\,\Om^{({\cal N})}_{i=2{\cal N}}
\ ,\quad D-1 \quad{\rm odd}\,,\nonumber
\eea
which in the cases of interest considered  above are given by the explicit expressions \re{SCS42}-\re{SCS53} with the terms $\om^{(D)}$ suppressed in both.
Such tasks are now under intensive consideration.

\bigskip
\noindent
{\bf Acknowledgments}
I am indebted to Eugen Radu and Francisco Navarro-L\'erida for extensive discussions and collaboration on aspects of this subject. Thanks also to Brian Dolan, Olaf Lechtenfeld,
Charles Nash, Denjoe O'Connor and Joost Slingerland for their comments.

\newpage
\appendix
\section{Higgs-Chern-Pontryagin (HCP) densities}
\setcounter{equation}{0}
\renewcommand{\theequation}{A.\arabic{equation}}
These are dimensional descendants of the Chern-Pontryagin (CP) densities in some higher (even) dimension. In the present context, the CP are defined as
the trace of an antisymmetrised product of anti-Hermitian $2{\cal N}\times 2{\cal N}$ Yang-Mills curvatures.
The bulk space is taken to be $M^D\times S^N$,
where $M^D$ is the residual space and $S^N$ the co-dimension. While $D+N=2{\cal N}$ is even, $D$ and $N$ can be both even, or, both odd.

The dimensional reduction scheme used is that introduced by Schwarz in \cite{Schwarz:1977ix},
and is applied in \cite{Schwarz:1981mb}.
The specific calculus of dimensional reduction employed here is an adaptation of the formalism in \cite{Schwarz:1981mb} to the case of codimension $S^N$ applied here.
This can be found in \cite{Tchrakian:2010ar} (and references therein). Basically, all we need here is what is given in
\cite{Tchrakian:2010ar}, the only difference being that we relax the restrictive choice of residual gauge groups imposed there. This can be made
transparent after stating the dimensional reduction formulas for the gauge connection ${\cal A}=({\cal A}_i,{\cal A}_I)$ on $M^D\times S^N$, with
$i=1,2,\dots,D$ and $I=1,2,\dots,N$.

In \cite{Schwarz:1977ix,Schwarz:1981mb}, the imposition of symmetry on the gauge connection ${\cal A}$ on $M^D\times K^N$ is carried out by solving the
symmetry equation involving the ingomogeneous term $g^{-1}\pa_Ig$, only at a fixed point of $K^N$.
At the fixed point the inhomogeneous term in the
symmetry equation is suppressed, rendering it an algebraic equation.
This algebraic equation encodes not the full group of invariance of $K^N$
but rather the stability (sub)group pertining to the fixed point. In the case of $S^N$ here, the symmetry groups are $SO(N+1)$ whose little groups are
$SO(N)$~\footnote{In the case of $S^2$ the little group is Abelian, this case being qualitatively different from all other $N\ge 3$ cases.}.

That the components $({\cal A}_i,{\cal A}_I)$ of the connection are given only at a fixed point  of $S^N$ is not a problem, since
the only quantities of interest, the action and the Chern-Pontryagin densities, are gauge and Lorentz invariant, being expressed in terms of the traces
of products of gauge covariant components $({\cal F}_{ij},{\cal F}_{iI},{\cal F}_{IJ})$ of the curvature.

An important consideration in the next two subsections will be the question of 
the choice of residual gauge group and Higgs multiplet which are two interrelated but separate issues. It is well known that after the descent,
the components of the gauge connection ${\cal A}_I$
on the codimension $K^N$ appear as scalars on the residual space $M^D$. For descents by $N\ge 2$, where the components of the curvature
${\cal F}_{IJ}$ do not vanish, an explicit symmetry breaking potential appears in the residual action density~\footnote{The action density in the
bulk space can be chosen to be any gauge invariant and Lorentz invariant density $\mbox{Tr}{\cal F}(2p)^2$, ${\cal F}(2p)$ being the
$p$-fold antisymmetrised product of the curvature ${\cal F}(2)$. For details, we refer to Ref. \cite{Tchrakian:2010ar}.}. Thus, the scalars on the
residual space $M^D$ can be described as Higgs fields. (When $N=1$, no explicit potential appears, but symmetry breaking asymptotics
necessary for finite energy, persists.)

\medskip
\noindent
{\bf The residual gauge group} must be chosen to be large enough for the Higgs-Chern-Simons (HCS) density to be nonvanishing. This is because the HCS in
$D-1$ dimensional (Minkowski) space is descended from the Higgs-Chern-Pontryagin (HCP) density in $D$ dimensions, and the latter is arrived at by
subjecting the ${\cal N}$-th Chern-Pontryagin (CP) density to dimensional descent down to $D$ dimensions, $i.e.$ ${\cal N}>D/2$. This means
that the residual gauge group must be large enough so that the trace of the ${\cal N}$-fold product of elements of its algebra do not vanish, 
leading to a straightforward group theoretical consideration in each case of interest.

\medskip
\noindent
{\bf The Higgs multiplet} in the residual field configuration is constrained by the requirement that the static solutions in the Yang-Mills--Higgs(YMH)
model on $\R^{D-2}$, before introducing the HCS term to the YMH Lagrangian, should support finite energy topologically stable solitons. Crudely, this means at least that the Higgs multiplets
must contain the adjoint representation of the residual gauge group and avoid being in the fundamental representation. (We will refine this
argument in Subsection {\bf A.2} below.) 

\medskip
In the next two subsections, we present the descent formulas for odd and even $N$ separately.
The technical reason for presenting odd and even $N$ cases separately
is, that the gamma matrices $\Ga_I$ ($i=1,2,\dots,N$) in $N$ dimensions are employed to represent the algebra of $SO(N)$, and it is only for even $N$
that there is a chiral matrix $\Ga_{N+1}$. As a result,
the descent formulas for even $N$ are qualitatively different from those for odd $N$.

\subsection{Descent over $S^N$: $N$ odd}
For the descent from the bulk dimension $2{\cal N}=D+N$ down to {\bf odd} $D$ (over odd $N$),
the components of the residual connection evaluated at the Noth pole of $S^N$ are given by
\bea
{\cal A}_i&=&A_i(\vec x)\otimes\eins\label{aiodd}\\
{\cal A}_I&=&\F(\vec x)\otimes\frac12\Gamma_I\,.\label{aIodd}
\eea
The unit matrix in \re{aiodd}, like the gamma matrix in $N$ dimensions in \re{aIodd}, are
$2^{\frac12(N-1)}\times 2^{\frac12(N-1)}$ arrays. If it is desired to have a traceless $n\times n$ anti-Hermitian
residual gauge connection $A_i(\vec x)$, then the
(traceless and anti-Hermitian) connection ${\cal A}_i$ in the bulk must be taken to be $2^{\frac12(N-1)}\,n\times 2^{\frac12(N-1)}\,n$.
The same goes for $\F$ and
${\cal A}_I$, except that now $\F$ does not have to be traceless~\footnote{In practice, when constructing soliton
solutions, $\F$ is taken to be traceless
without loss of generality.}

In \re{aiodd} and \re{aIodd}, and everywhere henceforth,
we have denoted the components of the residual coordinates as $x_i=\vec x$.
The dependence on the codimension coordinate $x_I$ is suppressed since all fields are evaluated at a fixed point
(North or South pole) of the codimension sphere.

The resulting components of the curvature are
\bea
{\cal F}_{ij}&=&F_{ij}(\vec x)\otimes\eins\label{fijodd}\\
{\cal F}_{iI}&=&D_{i}\F(\vec x)\otimes\frac12\Gamma_I\label{fiIodd}\\
{\cal F}_{IJ}&=&S(\vec x)\,\otimes\Gamma_{IJ}\,;\quad S\stackrel{\rm def.}=-(\eta^2\,\eins+\F^2)\,,\label{fIJodd}
\eea
where $\Gamma_{IJ}=-\frac14[\Gamma_{I},\Gamma_{J}]$ are the Dirac representation matrices of $SO(N)$ (the stability
group of the symmetry group of the $N$-sphere) and the dimensionful constant $\eta$ is the inverse of the radius of $S^N$,
which plays the role of the Higgs vacuum
expectation value.
In \re{fiIodd}, $D_{i}\F$ is the covariant derivative of the
Higgs field $\F$
\be
\label{covodd}
D_{i}\F=\pa_i\F+[A_i,\F]
\ee

\subsection{Descent over $S^N$:  $N$ even}
Because of the presence of chiral matrix $\Gamma_{N+1}$ in this case, the solution of the descent equations are
\bea
{\cal A}_i&=&A_i^{(+)}(\vec x)\otimes\,P_++A_i^{(-)}(\vec x)\otimes\,P_-
+\frac{i}{2}\,a_i(\vec x)\,\Gamma_{N+1}\label{aieven}\\
{\cal A}_I&=&\vf(\vec x)\otimes\frac12\,P_+\,\Gamma_I-\vf(\vec x)^{\dagger}\otimes\frac12\,P_-\,\Gamma_I\label{aIeven}\,,
\eea
where $\Gamma_{I}$ now are $2^{\frac{N}{2}}\times 2^{\frac{N}{2}}$, as are the projection operators
\be
\label{proj}
P_{\pm}=\frac12\left(\eins\pm\Gamma_{N+1}\right)\,.
\ee
Again, $A_i^{(\pm)}$ can be arranged to be $n\times n$ traceless anti-Hermitian connections, and $a_i$ is an Abelian factor. Unlike in the odd-$N$ case
however, the $n\times n$ complex arrays $\vf$ in \re{aIeven} are not constrained by Hermiticity of tracelessness.

The resulting components of the curvature are
\bea
{\cal F}_{ij}&=&F_{ij}^{(+)}(\vec x)\otimes\,P_++F_{ij}^{(-)}(\vec x)\otimes\,P_-
+\frac{i}{2}\ f_{ij}(\vec x)\,\Gamma_{N+1}\label{fijeven}\\
{\cal F}_{iI}&=&D_i\vf(\vec x)\otimes\,\frac12\,P_+\Gamma_I
-D_i\vf^{\dagger}(\vec x)\otimes\,\frac12\,P_-\Gamma_I\label{fiIeven}\\
{\cal F}_{IJ}&=&S^{(+)}(\vec x)\otimes\,P_+\Gamma_{IJ}+S^{(-)}(\vec x)\otimes\,P_-\Gamma_{IJ}\,,\label{fIJeven}
\eea
with the non-Abelian and Abelian curvatures
\bea
F_{ij}^{(\pm)}&=&\pa_iA_j^{(\pm)}-\pa_jA_i^{(\pm)}+[A_i^{(\pm)},A_j^{(\pm)}]\label{curvpm}\\
f_{ij}&=&\pa_ia_j-\pa_ja_i\,.\label{curvabel}
\eea
respectively, the covariant derivative
\bea
D_i\vf&=&\pa_i\vf+A_i^{(+)}\,\vf-\vf\,A_i^{(-)}+i\,a_i\,\vf\label{coveven}
\eea
and
\be
\label{Spm}
S^{(+)}=\vf\,\vf^{\dagger}-\eta^2\quad,\quad S^{(-)}=\vf^{\dagger}\,\vf-\eta^2\,.
\ee
The Abelian field can be absorbed by denoting $A_i^{(\pm)}$ formally as
\be
\label{redef}
\left(A_i^{(\pm)}\pm\frac{i}{2}\,a_i\,\eins\right)\rightarrow  A_i^{(\pm)}
\ee
Given the doubling of the two chiralities in \re{aieven}-\re{aIeven} and \re{fijeven}-\re{fIJeven}, all gauge invariant quantities can be
expressed in terms of the $(n\times n)\oplus (n\times n)$ chirally symmetric gauge connection and the Higgs field
\be
\label{chiral}
A_{i}\stackrel{\rm def.}=\left[
\begin{array}{cc}
A_{i}^{(+)} & 0\\
0 & A_{\mu}^{(-)}
\end{array}
\right]\ ,\quad
\Phi\stackrel{\rm def.}=\left[
\begin{array}{cc}
0 & \varphi\\
-\varphi^{\dagger} & 0
\end{array}
\right]
\ee
In this notation all gauge invariant quantities are expressed in terms of the gauge covariant quantities
\bea
F_{ij}&=&\pa_iA_j-\pa_jA_i+[A_i,A_j]=\left[
\begin{array}{cc}
F_{ij}^{(+)} & 0\\
0 & F_{ij}^{(-)}
\end{array}
\right]\label{Fijc}\\
D_{\mu}\Phi&=&\pa_i\F+[A_i,\F]=\left[
\begin{array}{cc}
0 & D_{i}\varphi\\
-D_{i}\varphi^{\dagger} & 0
\end{array}
\right]
\label{covPHY}\\
S&=&-(\eta^2\,\eins+\F^2)=\left[
\begin{array}{cc}
S^{(+)} & 0 \\
0 & S^{(-)}
\end{array}
\right]\,.\label{Sc}
\eea
Before proceeding to give some examples of Higgs--Chern-Pontryagin (HCP) densities, we point out that the above formulae
\re{aieven}-\re{aIeven} in the case of $N=2$ are not the most general. This is because, $SO(2)$, the little group of the symmetry group of $S^2$,
is Abelian in contrast to all other $S^N$, when this little group, $SO(N)$, is non-Abelian. As a result the (gauge) symmetry equations have (infinitely) more
numerous solutions as discussed in \cite{Schwarz:1981mb}. In this case \re{aieven}-\re{aIeven} are replaced by
\be
\label{N=2A}
{\cal A}_{i}=
A_{i}^{\al}\,\la_{\al} + a_{i}\,\la_{n(n+2)}\ ,\quad\al=1,2,\dots,n
\ee
where $(\la_{\al},\la_{n(n+2)})$ are the $SU(n)$ and $U(1)$ generators of the algebra of $SU(n+1)$, and,
\be
\label{fund12}
{\cal A}_{1}=\left[
\begin{array}{cccccc}
0 & 0 & 0 & 0 & 0 & \vf_1\\
0 & 0 & 0 & . & . & \vf_2\\
0 & 0 & 0 & . & . &  .\\
0 & . & . & . & . &   .\\
0 & . & . & . & . &   \vf_n\\
-\vf_1^{\dagger} & -\vf_2^{\dagger} & . & . & -\vf_n^{\dagger}  &  0
\end{array}
\right]\ ,\quad
{\cal A}_{2}=i\left[
\begin{array}{cccccc}
0 & 0 & 0 & 0 & 0 & \vf_1\\
0 & 0 & 0 & . & . & \vf_2\\
0 & 0 & 0 & . & . &  .\\
0 & . & . & . & . &   .\\
0 & . & . & . & . &   \vf_n\\
\vf_1^{\dagger} & \vf_2^{\dagger} & . & . & \vf_n^{\dagger}  &  0
\end{array}
\right]
\ee
where now the Higgs field in \re{fund12} consists of the fundemental multiplet
\be
\label{fund1}
\f=\left(
\begin{array}{c}
\vf_1\\
\vf_2\\
. \\
. \\
\vf_n
\end{array}
\right)
\ee
and the anti-fundamental $\f^{\dagger}$.

It can be checked that $({\cal A}_{i},{\cal A}_{I})$ for $N=2$ in \re{aieven}-\re{aIeven} are the special cases of
\re{N=2A}-\re{fund12} for $n=1$. Here, we have displayed
\re{fund12} only for completeness, and have ignored that option in Section {\bf 2}, in the construction of Higgs--Chern-Simons (HCS) densities. This is because no
finite energy soliton solutions for Higgs fields in this, \re{fund1}, representation can be constructed in any $\R^{D-2}$.

\subsection{Some examples of Higgs--Chern-Pontryagin (HCP) densities}
We restrict our attention here to HCP densities descended from the $3$-rd and $4$-th CP densities in $6$ and $8$ bulk dimensions. We further restrict
ourselves to residual dimensions $D\ge 4$ since HCP densities in lower dimensions than these are not relevant to the construction Higgs--Chern-Simons
densities. This is because $2+1$ dimensional Minkowski space is the lowest dimension in which a Chern-Simons density can be defined.


We now list the results of the application of the formulas \re{fijodd}-\re{fIJodd} and \re{fijeven}-\re{fIJeven}. The crucial property of the
residual HCP densities is that like the CP densities in the bulk, they are explicitly $total\ divergence$ as noted in \re{totdiv}. In the case of odd residual dimensions
we will use \re{fijodd}-\re{fIJodd} directly, but in the case of even residual dimensions we will use instead the notation given by
\re{chiral}, and, \re{Fijc}, \re{covPHY} and \re{Sc}, together with the $(n\times n)\oplus (n\times n)$ (formally) chiral matrix $\hat\Ga^{(n)}$
\be
\label{Gan}
\hat\Gamma^{(n)}=\left[
\begin{array}{cc}
-\eins_{n\times n} & 0_{n\times n}\\
0_{n\times n} & \eins_{n\times n}
\end{array}
\right]
\ee
In the list below, we give the explicit (total divergence) expressions for ${\cal C}_{\rm HCP}^{({\cal N},D)}=\bnabla\cdot\bOm^{({\cal N},D)}$ in the notation of \re{totdiv}.
For the HCP densities descended from the $3$-rd CP density we find the HCP densities ${\cal C}_{\rm HCP}^{(3,5)}$ and ${\cal C}_{\rm HCP}^{(3,4)}$
in $5$ and $4$ dimensional residual spaces, respectively,
\bea
{\cal C}_{\rm HCP}^{(3,5)}&=&\vep_{ijklm}\mbox{Tr}\ F_{ij}\,F_{kl}\,D_m\F
=\pa_m\Omega_m^{(3,5)}\,,\label{HCP35}\\
\Omega^{(3,5)}_m&=&\vep_{ijklm}\,\mbox{Tr}\ F_{ij}\,F_{kl}\,\F\label{Om35}
\eea
\bea
{\cal C}_{\rm HCP}^{(3,4)}&=&\vep_{ijkl}\,\mbox{Tr}\,\hat\Gamma^{(n)}\,\left(
S\,F_{ij}F_{kl}+2\,D_i\F\,D_j\F\,F_{kl}\right)
=\pa_i\Omega_i^{(3,4)}\label{HCP34}\\
\Omega^{(3,4)}_i&=&\vep_{ijkl}\mbox{Tr}\,\hat\Gamma^{(n)}\bigg[-2\eta^2A_{j}\left(F_{kl}
-\frac23\,A_{k}A_{l}\right)+\left(\F\, D_j\F-D_j\F\, \F\right)F_{kl}\bigg]\label{Om34}
\eea
For the HCP densities descended from the $4$-th CP density we find the HCP densities ${\cal C}_{\rm HCP}^{(4,7)}$, ${\cal C}_{\rm HCP}^{(4,6)}$,
${\cal C}_{\rm HCP}^{(4,5)}$ and ${\cal C}_{\rm HCP}^{(4,4)}$
in $7$, $6$, $5$ and $4$ dimensional residual spaces, in that order,
\bea
{\cal C}_{\rm HCP}^{(4,7)}
&=&\vep_{ijklmnp}\mbox{Tr}\ F_{ij}\,F_{kl}\,F_{mn}\,D_p\F=\pa_p\Omega_p^{(4,7)}\label{HCP47}\\
\Omega^{(4,7)}_p&=&\vep_{ijklmnp}\,\mbox{Tr}\ F_{ij}\,F_{kl}\,F_{mn}\,\F\label{Om47}
\eea
\bea
{\cal C}_{\rm HCP}^{(4,6)}
&=&\vep_{ijklmn}\,\mbox{Tr}\,\hat\Gamma^{(n)}\,
\bigg[S\,F_{ij}F_{kl}F_{mn}+2\,F_{ij}F_{kl}D_m\F D_n\F+F_{ij}D_m\F F_{kl}D_n\F\bigg]=\pa_i\Omega_i^{(4,6)}\label{HCP47}\\
\Omega^{(4,6)}_i&=&\vep_{ijklmn}\,\mbox{Tr}\,\hat\Gamma^{(n)}\,A_{j}\bigg[\left(F_{kl}F_{mn}-
F_{kl}A_{m}A_{n}+\frac25A_{k}A_{l}A_{m}A_{n}\right)+\nonumber\\
&&\hspace{40mm}+D_j\F\left(\F F_{kl}F_{mn}+F_{kl}\F F_{mn}+F_{kl}F_{mn}\F\right)\bigg]\label{Om46}
\eea
\bea
{\cal C}_{\rm HCP}^{(4,5)}&=&\vep_{ijklm}\,\mbox{Tr}
\bigg[ D_m\F\,(3\eta^2\,F_{ij}F_{kl}+F_{ij}F_{kl}\,\F^2+\F^2\,F_{ij}F_{kl}+F_{ij}\,\F^2 F_{kl})\nonumber\\
&&\hspace{60mm}-2F_{ij}D_k\F D_l\F D_m\bigg]=\pa_m\Omega_m^{(4,5)}\label{HCP45}\\
\Omega^{(4,5)}_m&=&\vep_{ijklm}\,\mbox{Tr}\bigg[
\F\left(\eta^2\,F_{ij}F_{kl}+\frac29\,\F^2\,F_{ij}F_{kl}+\frac19\,F_{ij}\F^2F_{kl}\right)
\nonumber\\
&&\qquad\qquad\qquad-\frac29\left(\F D_i\F D_j\F-D_i\F\F D_j\F+D_i\F D_j\F\F\right)F_{kl}\bigg]\label{Om45}
\eea
\bea
{\cal C}_{\rm HCP}^{(4,4)}&=&\vep_{ijkl}\,\mbox{Tr}\,\hat\Gamma^{(n)}\,\bigg[
2\,S^2\,F_{ij}F_{kl}+F_{ij}\,S\,F_{kl}\,S\nonumber\\
&&\hspace{30mm}+4\left(D_i\F\,D_j\F\,\{S,F_{kl}\}+D_i\F\,F_{kl}\,D_j\F\,S\right)\nonumber\\
&&\hspace{60mm}+2\,D_i\F\,D_j\F\,D_k\F\,D_l\F\bigg]=\pa_i\Omega_i^{(4,4)}\label{HCP44}\\
&&\nonumber\\
\Omega^{(4,4)}_i&=&\vep_{ijkl}\,\mbox{Tr}\,\hat\Gamma^{(n)}\,\bigg\{6\eta^4\,A_j\left(F_{kl}-\frac23\,A_k\,A_l\right)
-6\,\eta^2\left(\F\,D_j\F-D_j\F\,\F\right)\,F_{kl}\nonumber\\
&&\hspace{20mm}+\bigg[\left(\F^2\,D_j\F\,\F-\F\,D_j\F\,\F^2\right)-2\left(\F^3\,D_j\F-D_j\F\,\F^3\right)\bigg]F_{kl}
\bigg\}\label{Om44}
\eea
The most important property of the above displayed residual CP densities $D$ dimensions, namely the HCP densities ${\cal C}_{\rm HCP}^{({\cal N},D)}$ \re{HCP35}-\re{HCP44}, 
is that they are $total\ divergence$~\footnote{We note in passing that this all-important property of ``total divergence'' persists also for the descents \re{N=2A}-\re{fund12}
involving fundamental Higgs multiplets,
which for physical reasons are ignored here. The analogue of \re{HCP34} in this case can be found in \cite{Tchrakian:1985au}.}.

For ${\cal N}=3$ and ${\cal N}=4$, these are listed above. The residual gauge connection in odd $D$ is the
anti-Hermitian $n\times n$ matrix in \re{aiodd}, while the residual gauge connection in even $D$ is the anti-Hermitian $(n\times n)\oplus (n\times n)$
matrix \re{chiral}. The integer $n$ in each case can be chosen according to physical requirements.


\medskip
\noindent
{\bf A special case of interest} is the case $n=1$, $i.e.$ when the residual gauge group is Abelian. This is the case when
the effect of adding the Higgs--Chern-Simons (HCS) term to the Lagrangian on the solitons of Abelian Higgs systems is to be studied, $e.g.$ in the case of
Abelian-Higgs vortices in $2+1$ dimensions. (Note that the usual Chern-Simons term is the leading term in the HCS density.)
Another area of potential interest is in the study black holes interacting with the Maxwell field $e.g.$ in $4+1$ dimensions.

It is clear from \re{aiodd} that for odd $D$ there is no possibility of sustaining a residual Abelian gauge field, while this is possible for even $D$ as seen
from \re{aieven}. In the latter case, to construct the HCP density of an Abelian system, one can fix $n=1$ in \re{aieven} and suppress the non-Abelian quantities $A_i^{(\pm)}$
retaining only the Abelian connection $a_i$ and the complex scalar $\vf$, so that \re{aieven}-\re{aIeven} is now replaced by
\bea
{\cal A}_i&=&\frac{i}{2}\,a_i(\vec x)\,\Gamma_{N+1}\label{aiabel}\\
{\cal A}_I&=&\vf(\vec x)\ \frac12\,P_+\,\Gamma_I-\vf(\vec x)^{*}\ \frac12\,P_-\,\Gamma_I\label{aIabel}\,,
\eea
It turns out that the HCP density in this case vanishes in $D=4p$ for $p=1,2,\dots$ and is nonvanishing in $D=4p+2$, for $p=0,1,2,\dots$.
To distinguish these from \re{HCP35}-\re{HCP44}, we denote them instead as $\tilde{\cal C}_{\rm HCP}^{({\cal N},D)}$, in which notation all
\[
\tilde{\cal C}_{\rm HCP}^{({\cal N},D=4p)}=0 \ ,\quad p=1,2,\dots
\]
vanish. Concerning the nonvanishing Abelian HCP densities, we apply the same restriction~\footnote{Although HCP densities in $D=2$ (the $p=0$ case in $D=4p+2$)
are irrelevant to the construction of Chern-Simons densities,
it may be interesting to see that they indeed do not vanish and are also total divergence. The $N=4$ and $N=6$ examples are
\bea
\tilde{\cal C}_{\rm HCP}^{(3,2)}&=&\vep_{ij}\pa_i\left[i\eta^4\,a_j+(|\vf|^2-2\eta^2)\vf^*D_j\vf\right]\label{D=2,1}\\
\tilde{\cal C}_{\rm HCP}^{(4,2)}&=&\vep_{ij}\pa_i\left[-i\eta^6\,a_j+\left((|\vf|^2)^2-3\eta^2|\vf|^2+3\eta^4\right)\vf^*D_j\vf\right]\label{D=2,2}
\eea} as in \re{HCP35}-\re{HCP44} above, since only HCP densities
in dimensions $D\ge 4$ are relevant to the construction of Chern-Simons densities.
For simplicity, we also restrict our attention to descents of $N=2$, $i.e.$ ${\cal N}=2(p+1)$. In this case, $\tilde{\cal C}_{\rm HCP}^{({\cal N},D}=\tilde{\cal C}_{\rm HCP}^{({\cal N}=2p+2,D=4p+2)}$, can
be expressed compactly for the arbitrary $p$ case as
\be
\label{calN=2p+2,D=4p+2}
\tilde{\cal C}_{\rm HCP}^{(2p+2,4p+2)}
=\vep_{i_1i_2\dots i_{4p+1}i_{4p+2}}\pa_{i_{4p+2}}\left[f_{i_1i_2}f_{i_3i_4}\dots f_{i_{4p-1}i_{4p}}\left(\eta^2\,a_{i_{4p+1}}+i\vf^*D_{i_{4p+1}}\vf\right)\right]
\ee
which is manifestly $total\ divergence$.

\section{Skyrme-Chern-Pontryagin (SCP) densities}
\setcounter{equation}{0}
\renewcommand{\theequation}{B.\arabic{equation}}
The quantities referred to here as Skyrme-Chern-Pontryagin (SCP) densities are nothing else than the topological charge densities of
$SO(N)$ gauged $O(D+1)$ Skyrme
scalars, where~\footnote{This integer $N$ here should not be confused with that appearing in $S^N$ in Appendix {\bf A}.
It is quite distinct from the latter, but
plays a curiously analogous role.} $2\le N\le D$. By construction, these are not dimensional
descents of Chern-Pontryagin (CP) densities. What they have in common with CP
densities is that they are topological charge densities in $\R^D$, and like the latter are also
(essentially) $total\ divergence$. The nomenclature containing ``Chern-Pontryagin''
is chosen to underline the analogy with Higgs--Chern-Pontryagin densities discussed above.

The prescription for constructing the topological charge density for the $SO(N)$ gauged $O(D+1)$ (Skyrme)
sigma model ($2\le N\le D$) on $\R^D$ is presented in this Appendix.
This is mainly the work in Ref. \cite{LMP}, adapted to the purposes of the present report. Here, we are
not concerned with the topological charge of a given Skyrme model $per\ se$~\footnote{Ultimately the objective is
to study the effect that the new Chern-Simons has on the soliton on $\R^{D-2}$, so in this respect the
topological charge is only of academic interest here.}. Once we have defined the charge density in
$D$ dimensions, on some space $M^D$, all that interests us is its
$total\ divergence$ expression from which we extract the definition of the corresponding
Chern-Simons density (SCS) in $M^{D-1}$.

In the following subsections, we first recall the definition of the topological charge, or winding number, density of the
(ungauged) Skyrmion in $\R^D$. Next we present the prescription of
constructing the topological charge density for the $SO(N)$ gauged Skyrmion in $\R^D$, and finally we list
some relevant examples of these densities which are referred to as
Skyrme--Chern-Pontryagin (SCP) densities.

\subsection{Winding number density of Skyrmions on $\R^D$}
Skyrme sigma models are the $O(D+1)$ sigma models in {\bf all} $D$-dimensions
They are defined in terms of a scalar field $\f^a$, $a=1,2,\dots,D+1$ subject to the constraint
\be
\label{constr}
|\f^a|^2=1
\ee
The Lorentz invariant density
\be
\label{wN1}
\varrho_0^{(D)} =\vep_{i_1 i_2 ...i_{D}} \vep^{a_1 a_2 ...a_D a_{D+1}}\pa_{i_1} \phi^{a_1}\
 \pa_{i_2} \phi^{a_2} ...\pa_{i_D}\phi^{a_{D}} \: \phi^{a_{D+1}}
\ee
is $essentially\ total\ divergence$ in the sense that when subjected to the variational principle, with the constraint \re{constr} taken into account,
there result no equations of motion.

When the $D$-dimensional space $M^D$ is $\R^{D}$, then the volume integral of $\varrho_0^{(D)}$ is a $winding\ number$ (up to angular volume),
$i.e.$ it is an $integer$ qualifying it as a topological charge density.

If one chooses to parametrise $\f^a$ with the 'polar' functions $f^{(1)},\,f^{(2)},\dots,f^{(D-1)}$, and the 'azimuthal' function $g$ on $S^{D}$, one has
\bea
\varrho_0^{(D)}
&=&\vep_{i_1 i_2 ...i_{D}}\,\left(\sin^{D-1}f^{(1)}\pa_{i_1}f^{(1)}\right)\left(\sin^{D-2}f^{(2)}\pa_{i_2}f^{(2)}\right)
\dots\left(\sin f\pa_{i_{D-1}}f^{(D-1)}\right)\pa_{i_{D}}g\label{wN2}\\
&=&\pa_{i_{D}}\Bigg(\vep_{i_1 i_2 ...i_{D}}\,\left(\sin^{D-1}f^{(1)}\pa_{i_1}f^{(1)}\right)\left(\sin^{D-2}f^{(2)}\pa_{i_2}f^{(2)}\right)
\dots\left(\sin f\pa_{i_{D-1}}f^{(D-1)}\right)g\Bigg)\label{wN3}
\eea
which is manifestly a $total\ divergence$. Of course the total divergence property \re{wN3} is not unique, and any other one of the other $D-1$
partial derivatives could have been privileged. This choice here is made merely on typographical convenience.

We express \re{wN3} as
\be
\label{denote}
\varrho_0^{(D)}\stackrel{\rm def.}=\pa_{i_D}\omega_{i_D}^{(D)}\,,
\ee
in a simplified notation.


\subsection{Topological charge density of $SO(N)$ gauged Skyrmions}
Dividing the components fof the Skyrme scalar as $\f^a=(\f^{\al},\f^A)$ with $\al=1,2,\dots,N$ and $A=N+1,N+2,\dots,D+1$, we proceed to
gauge $N$ components $\f^{\al}$ with $SO(N)$ and leave the components $\f^A$
\bea
D_i\f^{\al}&=&\pa_i\f^{\al}+A_i\f^{\al}\label{Dal}\\
D_i\f^A&=&\pa_i\f^{A}\label{DA}
\eea
$A_i$ being the $SO(N)$ gauge connection $A_i^{\al\beta}=-A_i^{\beta\al}$. In \re{Dal} and henceforth we use the notation
\be
\label{notation}
A_i\f^{\al}=A_i^{\al\beta}\f^{\beta}\,.
\ee
It is clear that the topological charge density $\varrho_0^{(D)}$, \re{wN1}, while it is
(Lorentz invariant and) total divergence, it is not invariant under the gauge transformation \re{Dal}-\re{DA}.

One can construct the Lorentz invariant density for the Skyrme scalar $\f^a$
\be
\label{rG}
\varrho_N^{(D)} =\vep_{i_1 i_2 ...i_{D}} \vep^{a_1 a_2 ...a_D a_{D+1}}D_{i_1}\phi^{a_1}D_{i_2} \phi^{a_2}\dots D_{i_D}\phi^{a_{D}} \: \phi^{a_{D+1}}
\ee
which is invariant under $SO(N)$ gauge transformations, but unlike $\varrho_0^{(D)}$ is {\bf not} $total\ divergence$.

The required "topological charge" density $\varrho^{(D)}$ must be Lorentz invariant, gauge invariant, {\bf and} be $total\ divergence$.
To this end, one casts the difference between $\varrho_N^{(D)}$ and $\varrho_0^{(D)}$ in the form
\be
\label{Om-W}
\varrho_N^{(D)}-\varrho_0^{(D)}=\bnabla\cdot\bOm^{(D,N)}[A,\f]-W^{(D,N)}[F,D\f]\,.
\ee
The superscript $(D,N)$ on $W$ and $\bOm$ indicate that the Skyrme model is in $D$ dimensions, and it is gauged with $SO(N)$.
The scalar $W^{(D,N)}[F,D\f]$ is expressed in terms of the curvature $F$, and the covariant derivative $D\f$, and is therefore gauge invariant.
The quantity $\bOm^{(D,N)}[A,\f]\equiv\Omega^{(D,N)}_i[A,\f]$ however is expressed in terms of the connection $A$, and is gauge variant.

Rearranging \re{Om-W} leads to the two equivalent definitions of the topological charge density $\varrho^{(D,N)}$,
\bea
\varrho^{(D,N)}&\stackrel{\rm def.}=&\varrho_N^{(D)}+W^{(D,N)}[F,D\f]\label{gaugeinv}\\
&\stackrel{\rm def.}=&\varrho_0^{(D)}+\bnabla\cdot\bOm^{(D,N)}[A,\f]\label{totdivs}
\eea
Note that the quantity $\bOm^{(D,N)}$ in \re{totdivs} is distinct from its Higgs analogue $\bOm^{({\cal N},D)}$ appearing in \re{totdiv}.

Both terms in \re{gaugeinv} are separately gauge invariant~\footnote{In the context
of establishing lower bounds on the energy of static fields on $\R^D$, this
expression of the topologcal charge is used in the Bogomol'nyi analysis.}, while definition \re{totdivs}
consists of two gauge variant terms whose sum is of course
gauge invariant. The advantage \re{totdivs} has is that it is expressed a total divergence explicitly,
which is the property needed for the definition of Chern-Simons densities.

For $\varrho^{(D,N)}$ to qualify as a topological charge density, its volume integral in $\R^D$ must be an integer (up to a factor of the angular
volume). This is the case with the volume integral of $\varrho_0^{(D)}$ in \re{totdivs}, as defined by \re{wN1}. It follows that the surface
integral of the term
\[
\bOm^{(D,N)}[A,\f]
\]
in \re{totdivs} must vanish. This can always be arranged in any given example, but we are not concerned with that question here.

What interests us here is, that in \re{totdivs} we have an expression for the topological charge density $\varrho^{(D,N)}$ which is explicitly
total divergence and is the Skyrme-analogue of the usual CP density. This is what we referred to above as the Skyrme--Chern-Pontryagin (SCP)
density ${\cal C}_{\rm SCP}^{(D,N)}$. Identifying ${\cal C}_{\rm SCP}^{(D,N)}$ with the topological density $\varrho^{(D,N)}$, we rewrite
the useful expression \re{totdivs} as
\bea
\label{SCP}
{\cal C}_{\rm SCP}^{(D,N)}&=&\varrho_0^{(D)}+\bnabla\cdot\bOm^{(D,N)}[A,\f]\nonumber\\
&=&\bnabla\cdot\left(\bom^{(D)}+\bOm^{(D,N)}\right)\,,
\eea
having used the notation in \re{denote}. Several instructive examples of \re{gaugeinv}-\re{totdivs} are given in the next subsection.

We now note that in all even dimensions, say $D=2{\cal N}$, in addition to the SCP densities \re{SCP}, there exist also the $usual$ ${\cal N}$-th CP densities
${\cal C}_{\rm CP}^{({\cal N})}$, \re{usualtt}.
These (usual) CP densities can be added to the SCP densities \re{gaugeinv}-\re{totdivs}, while
retaining the all-important total divergence property.
Since in any given dimension the gauging prescription \re{Dal}-\re{DA} provides for various groups $SO(N)$ with $2\le N\le D$,
this information is implicitly encoded in the relevant ${\cal C}_{\rm CP}^{({\cal N})}$, $i.e.$ that the ${\cal N}$-th CP term in question is that for gauge group $SO(N)$.

Thus in even dimensions the definition of the SCP density \re{SCP} is extended by the addition of the usual CS term in those dimensions,
for the given gauge group. To make this clear, we introduce the modified notation
\bea
\label{SCPe}
\hat{\cal C}_{\rm SCP}^{(D,N)}&=&{\cal C}_{\rm SCP}^{(D,N)}+\la\,{\cal C}_{\rm CP}^{({\cal N})}\nonumber\\
&=&\bnabla\cdot\left(\bom^{(D)}+\bOm^{(D,N)}\right)+\la\,{\cal C}_{\rm CP}^{({\cal N})}\,,
\eea
where $\la$ is some real number to be precised in the next subsection.

\subsection{Some examples of Skyrme--Chern-Pontryagin (SCP) densities}
We will list some instructive (but not exclusively relevant to the task here) examples of \re{gaugeinv}-\re{totdivs} SCP densities here, in three groupings. The first of these
pertains to gauging with $SO(D)$ for $D=2,3,4$. The second one is for gauge group $SO(3)$ for $D=4,5$ and the third
for gauge group $SO(2)$ for $D=3,4,5$. This way one can glean some general features of the SCP densities.

While in Section {\bf 3} we will need only \re{totdivs} , or \re{SCP} in the construction of Skyrme--Chern-Simons (SCS) densities, here we give also
the versions given by \re{gaugeinv} as an illustration of the results.

The index notation used is that in \re{Dal}-\re{DA}, the coordinates on the $D$-dimensional space being $i=1,2,\dots,D$,
$\al=1,2,\dots,N$ and $A=N+1,N+2,\dots,D+1$.

\subsubsection{SCP densities for $SO(D)$ gauging: $N=D=2,3,4$}
$N=D=2$, $SO(2)$: $i=1,2$, $\al=1,2$ and $A=3$
\bea
{\cal C}_{\rm SCP}^{(2,2)}&=&\varrho_2^{(2)}+\frac12\,\vep_{ij}\vep^{\al\al'} \phi^3 F_{ij}^{\al\al'}\equiv\varrho_2^{(2)}+\vep_{ij} \phi^3 F_{ij}\label{SCP22a}\\
&=&\pa_{i}\left(\om_i^{(2)}+\vep_{ij}\vep^{\al\al'}\phi^3A_{j}^{\al\al'}\right)\equiv\pa_{i}\left(\om_i^{(2)}+2\,\vep_{ij}\phi^3A_{j}\right)\ ,\quad \vep^{\al\al'}A_{j}^{\al\al'}=2A_i\label{SCP22b}
\eea
$N=D=3$, $SO(3)$: $i=1,2,3$, $\al=1,2,3$ and $A=4$
\bea
{\cal C}_{\rm SCP}^{(3,3)}
&=&\varrho_3^{(3)}-\frac{3}{2}\,\vep_{ijk}\vep^{\alpha \beta \beta'}\phi^{\alpha} F_{ij}^{\beta \beta'} D_{k}\phi^4\label{SCP33a}\\
&=&\pa_{k}\left(\om_k^{(3)}+3\,\vep_{ijk}\vep^{\alpha \beta \beta'}\phi^{\alpha}\pa_{i} \phi^4 A_j^{\beta \beta'}\right)\label{SCP33a}
\eea$
N=D=4$, $SO(4)$: $i=1,2,3,4$, $\al=1,2,3,4$ and $A=5$
\bea
{\cal C}_{\rm SCP}^{(4,4)}&=&\varrho_4^{(4)}+\vep_{ijkl}\vep^{\alpha \beta \gamma \delta} \phi^5
[3F_{kl}^{\gamma \delta} \: D_{i} \phi^{\alpha} D_{j}
\phi^{ \beta} + \frac{1}{4} (\phi^5)^2 F_{ij}^{\alpha \beta} F_{kl}^{ \gamma \delta}]\label{SCP44a}\\
&=&\pa_{i}\Bigg(\om_i^{(4)}+\frac32\vep_{ijkl}\vep^{\alpha \alpha'\beta\beta'}\left[
\phi^5 (1-\frac{1}{3} (\phi^5)^2) A_{j}^{\al\al'}
\left(\pa_{k} A_{l}^{\beta\beta'}+\frac{2}{3} (A_{k} A_{l} )^{\beta \beta'}\right)\right] \nonumber\\
&&\hspace{20mm}+3\,\f^5
\bigg[F_{jk}^{\al\al'}\f^{\beta}D_l\f^{\beta'}-\pa_j\left[A_k^{\al\al'}\f^{\beta}\left(2\pa_l\f^{\beta'}+A_l\f^{\beta'}\right)\right]\bigg]\Bigg)\label{SCP44b}
\eea
We can now reconsider the extended definition \re{SCPe} in the case of even $D$.
For the examples \re{SCP22a}-\re{SCP22b} and \re{SCP44a}-\re{SCP44b} at hand, the ${\cal C}_{\rm CP}^{({\cal N})}$ in
question are the (usual) $1$-st and $2$-nd CP invariants
with gauge groups $G=SO(2)$ and $G=SO(4)$ respectively
\bea
{\cal C}_{\rm CP}^{(1)}&=&\frac12\,\vep_{ij}\,\vep_{\al\al'}\,F_{ij}^{\al\al'}=\vep_{ij}\,\vep_{\al\al'}\pa_iA_j^{\al\al'}\label{CP1}\\
{\cal C}_{\rm CP}^{(2)}&=&\frac14\,\vep_{ijkl}\,\vep_{\al\al'\beta\beta'}\,F_{ij}^{\al\al'}F_{kl}^{\beta\beta'}\nonumber\\
&=&\pa_i\,\vep_{ijkl}\,\vep_{\al\al'\beta\beta'}A_{j}^{\al\al'}
\left(\pa_{k} A_{l}^{\beta\beta'}+\frac{2}{3} (A_{k} A_{l} )^{\beta \beta'}\right)\label{CP2}
\eea
For $\la=-1$ the volume integral of $\hat{\cal C}_{\rm SCP}^{(2,2)}$ is equal to the volume integral of
$\vr_0^{(2)}=\pa_i\om_i^{(2)}$, namely to the winding number,
which is what is required to qualify it as a topological charge density~\footnote{Subtracting \re{CP1} from \re{SCP22b} yields
\[
{\cal C}_{\rm SCP}^{(2,2)}
=\pa_{i}\left(\om_i^{(2)}+\vep_{ij}\vep^{\al\al'}(\phi^3-1)A_{j}^{\al\al'}\right)\,.
\]
The volume integral of the second term does has vanishing contribution. This is because it is evaluated as a surface integral and $\f^3=1$ asymptotically
by virtue of finite energy conditions at infinity. Thus only the first term contributes, namely the winding number.

Likewise, subtracting \re{CP2} from \re{SCP44b} yields
\bea
{\cal C}_{\rm SCP}^{(4,4)}
&=&\pa_{i}\Bigg(\om_i^{(4)}+\frac32 \vep_{ijkl}\vep^{\alpha \alpha'\beta\beta'}\bigg[
\left(-\frac23+\phi^5 (1-\frac{1}{3} (\phi^5)^2)\right) A_{j}^{\al\al'}
\left(\pa_{k} A_{l}^{\beta\beta'}+\frac{2}{3} (A_{k} A_{l} )^{\beta \beta'}\right) \nonumber\\
&&\hspace{20mm}+3\,\f^5
\bigg[F_{jk}^{\al\al'}\f^{\beta}D_l\f^{\beta'}-\pa_j\left[A_k^{\al\al'}\f^{\beta}\left(2\pa_l\f^{\beta'}+A_l\f^{\beta'}\right)\right]\bigg]\Bigg)\nonumber
\eea
Again the contribution of the volume integral is only the winding number since the integrand in the relevant surface integral vanishes. The first two terms there vanish due to
$\f^5=1$ at infinity, and the last term vanishes since $\f^5=1$ implies $|\f^{\al}|=0$.

The effects of these two subtractions on \re{SCP22a} and \re{SCP44a} are also interesting. The results are
\bea
{\cal C}_{\rm SCP}^{(2,2)}&=&\varrho_2^{(2)}+\frac12\,\vep_{ij}\vep^{\al\al'} (\phi^3-1) F_{ij}^{\al\al'}\,,\nonumber\\
{\cal C}_{\rm SCP}^{(4,4)}&=&\varrho_4^{(4)}+\vep_{ijkl}\vep^{\alpha \beta \gamma \delta} 
\left[3\f^5F_{kl}^{\gamma \delta} \: D_{i} \phi^{\alpha} D_{j}
\phi^{ \beta} + \frac{1}{4} \left((\phi^5)^3-1\right) F_{ij}^{\alpha \beta} F_{kl}^{ \gamma \delta}\right]\,.\nonumber
\eea
In the corresponding Bogomol'nyi analyses, respectively, the two ``pion-mass'' potentials
\[
(\f^3-1)^2\quad{\rm and}\quad ((\f^5)^3-1)^2
\]
emerge naturally. The first of these ``pion-mass'' potentials, namely that in $2+1$ dimensions, appears in \cite{Schroers:1995he}, where
the charge density \re{SCP22a}-\re{SCP22b} is also given in a slightly different formulation.}. While this is a very important property in the context of solitons, in the
subsequent step of defining a Skyrme--Chern-Simons density choosing $\la=-1$ is not of special importance.

\subsubsection{SCP densities for $SO(3)$ gauging: $D=4,5$}
$N=3$, $D=4$, $SO(3)$: $i=1,2,3,4$, $\al=1,2,3$ and $A=4,5$
\bea
{\cal C}_{\rm SCP}^{(4,3)}&=&\vr_3^{(4)}+2\,\vep_{{i}{j}{k}{l}} \vep^{{A}{B}}
F_{{i}{j}}^{\al}\,\f^{\al}D_k\,\f^{{A}}\,D_l\,\f^{{B}}\label{SCP43a}\\
&=&\pa_{{i}}\left(\om_i^{(4)}+4\,\vep_{{i}{j}{k}{l}} \vep^{{A}{B}}
A_{{j}}^{\al}\,\f^{\al}\pa_{{k}}\,\f^{{A}}\,\pa_{{l}}\,\f^{{B}}\right)\label{SCP43b}
\eea
$N=3$, $D=5$, $SO(3)$: $i=1,2,3,4,5$, $\al=1,2,3$ and $A=4,5,6$
\bea
{\cal C}_{\rm SCP}^{(5,3)}&=&\vr_3^{(5)}
+\frac12\,\vep_{{i}{j}{k}{l}{m}}\vep^{{A}{B}{C}}F_{{i}{j}}^{\al}\,\f^{\al}
D_k\f^{{A}}D_l\f^{{B}}D_m\f^{{C}}\label{SCP53a}\\
&=&\pa_{{i}}\left(\om_i^{(5)}+\vep_{{i}{j}{k}{l}{m}}\vep^{{A}{B}{C}}\,A_{{j}}^{\al}\,\f^{\al}
\pa_{{k}}\f^{{A}}\pa_{{l}}\f^{{B}}\pa_{{m}}\f^{{C}}\right)\label{SCP53b}
\eea
In the $4$-dimensional case \re{SCP43a}-\re{SCP43b}, one can add/subtract the $2$-nd $SO(3)$ CP density
\bea
{\cal C}_{\rm CP}^{(2)}&=&\vep_{ijkl}\,F_{ij}^{\al}\,F_{kl}^{\al}\nonumber\\
&=&4\,\vep_{ijkl}\,A_{j}^{\al}
\left(\pa_{k} A_{l}^{\al}+\frac{2}{3} (A_{k} A_{l} )^{\al}\right)\label{CP21}
\eea

\subsubsection{SCP densities for $SO(2)$ gauging: $D=3,4,5$}
$N=2$, $D=3$, $SO(2)$: $i=1,2,3$, $\al=1,2$ and $A=3,4$
\bea
{\cal C}_{\rm SCP}^{(3,2)}&=&\varrho_2^{(3)} +\vep_{ijk} \vep^{AB} F_{ij}\,D_k\phi^A\phi^B \label{SCP32a}\\
&=&\pa_i\left(\om_i^{(3)} +\vep_{ijk}\vep^{AB}  A_j\, \pa_k \phi^A \, \phi^B\right)\label{SCP32b}
\eea
$N=2$, $D=4$, $SO(2)$: $i=1,2,3,4$, $\al=1,2$ and $A=3,4,5$
\bea
{\cal C}_{\rm SCP}^{(4,2)}&=&\vr_2^{(4)}+\vep_{{i}{j}{k}{l}} \vep^{{A}{B}{C}}
F_{{i}{j}}\,D_{{k}}\,\f^{{A}}\,D_{{l}}\,\f^{{B}}\,\f^{{C}}\label{SCP42a}\\
&=&\pa_{{i}}\left(\om_i^{(4)}
+2\,\vep_{{i}{j}{k}{l}} \vep^{{A}{B}{C}}\,A_{{j}}\,\pa_{k}\,\f^{{A}}\,\pa_{{l}}\,\f^{{B}}\,\f^{{C}}\right)\label{SCP42b}
\eea
$N=2$, $D=5$, $SO(2)$: $i=1,2,3,4,5$, $\al=1,2$ and $A=3,4,5,6$
\bea
{\cal C}_{\rm SCP}^{(5,2)}&=&\vr_2^{(5)}+\vep_{{i}{j}{k}{l}{m}}\vep^{{A}{B}{C}{D}}F_{{i}{j}}\,
D_{{k}}\f^{{A}}D_{{l}}\f^{{B}}D_{{m}}\f^{{C}}\f^{{D}}\label{SCP52a}\\
&=&\pa_{{i}}\left(\om_i^{(5)}+2\,\vep_{{i}{j}{k}{l}{m}}\vep^{{A}{B}{C}{D}}\,A_{{j}}\,
\pa_{{k}}\f^{{A}}\pa_{{l}}\f^{{B}}\pa_{{m}}\f^{{C}}\f^{{D}}\right)\label{SCP52b}
\eea
Again, in the $4$-dimensional case \re{SCP42a}-\re{SCP42b}, one can add/subtract the $2$-nd Abelian CP density
\be
\label{CPabel}
{\cal C}_{\rm CP}^{(2)}=\vep_{ijkl}\,F_{ij}F_{kl}=2\,\vep_{ijkl}\,\pa_i\left(A_j\,F_{kl}\right)
\ee

\newpage
\begin{small}

\end{small}


\begin{thebibliography}{99}
\bibitem{Jackiw:1985}
see for example, R.~Jackiw, "Chern-Simons terms and cocycles in physics and mathematics",
in E.S. Fradkin $Festschrift$, Adam Hilger, Bristol (1985).
\bibitem{Deser:1982vy}
  S.~Deser, R.~Jackiw and S.~Templeton,
  Phys.\ Rev.\ Lett.\  {\bf 48} (1982) 975.
\bibitem{Deser:1981wh}
  S.~Deser, R.~Jackiw and S.~Templeton,
  Annals Phys.\  {\bf 140} (1982) 372
   [Annals Phys.\  {\bf 185} (1988) 406]
   [Annals Phys.\  {\bf 281} (2000) 409].
\bibitem{Paul:1986ix}
  S.~K.~Paul and A.~Khare,
  Phys.\ Lett.\ B {\bf 174} (1986) 420
   [Erratum-ibid.\  {\bf 177B} (1986) 453].
\bibitem{Hong:1990yh}
  J.~Hong, Y.~Kim and P.~Y.~Pac,
  Phys.\ Rev.\ Lett.\  {\bf 64} (1990) 2230.
\bibitem{Jackiw:1990aw}
  R.~Jackiw and E.~J.~Weinberg,
  Phys.\ Rev.\ Lett.\  {\bf 64} (1990) 2234.
\bibitem{Ghosh:1995ze}
  P.~K.~Ghosh and S.~K.~Ghosh,
  Phys.\ Lett.\ B {\bf 366} (1996) 199
  [hep-th/9507015].
\bibitem{Kimm:1995mi}
  K.~Kimm, K.~-M.~Lee and T.~Lee,
  Phys.\ Rev.\ D {\bf 53} (1996) 4436
  [hep-th/9510141].
\bibitem{Arthur:1996uu}
  K.~Arthur, D.~H.~Tchrakian and Y.~-s.~Yang,
  Phys.\ Rev.\ D {\bf 54} (1996) 5245.
\bibitem{Schwarz:1977ix}
  A.~S.~Schwarz,
  Commun.\ Math.\ Phys.\  {\bf 56} (1977) 79.
\bibitem{Schwarz:1981mb}
  A.~S.~Schwarz and Y.~S.~Tyupkin,
  Nucl.\ Phys.\ B {\bf 187} (1981) 321.
\bibitem{Tchrakian:2010ar}
 D.~H.~Tchrakian,
  J.\ Phys.\ A {\bf 44} (2011) 343001
  [arXiv:1009.3790 [hep-th]].
\bibitem{Radu:2011zy}
  E.~Radu and T.~Tchrakian,
in Proceedings of the XV International Conference on Symmetry
Methods in Physics,
(Dubna, Russia, July 12-16, 2011 and Yerevan, July 25-29, 2011) Journal of
Physics of Atomic Nuclei {\bf 76}(10), 2013.
  arXiv:1101.5068 [hep-th].
\bibitem{Navarro-Lerida:2013pua}
  F.~Navarro-L\'erida, E. Radu, and D.~H.~Tchrakian,
  Int. J. Mod. Phys. A {\bf 29} (2014) 1450149
  [arXiv:1311.3950 [hep-th]].
\bibitem{Navarro-Lerida:2014rwa}
  F.~Navarro-Lerida and D.~H.~Tchrakian,
  Int.\ J.\ Mod.\ Phys.\ A {\bf 30} (2015) 15,  1550079
  [arXiv:1412.4654 [hep-th]].
\bibitem{Skyrme:1962vh}
  T.~H.~R.~Skyrme,
  Nucl.\ Phys.\  {\bf 31} (1962) 556.
\bibitem{LMP}
 D.~H.~Tchrakian,
 Lett. Math. Phys. {\bf 40} (1997) 191-201.
\bibitem{Schroers:1995he}
  B.~J.~Schroers,
  Phys.\ Lett.\ B {\bf 356} (1995) 291
  [hep-th/9506004].
\bibitem{Julia:1975ff}
B.~Julia and A.~Zee,
  Phys.\ Rev.\ D {\bf 11} (1975) 2227.
\bibitem{Tchrakian:1985au}
  D.~H.~Tchrakian,
  Phys.\ Lett.\ B {\bf 155} (1985) 255.

  

\end{thebibliography}
\end{document}